\documentclass[aps,prd,reprint,groupedaddress,nofootinbib]{revtex4-2}

\usepackage{amsmath}
\usepackage{booktabs}
\usepackage{longtable}
\usepackage{threeparttablex}
\usepackage{graphicx}
\usepackage{multirow}
\usepackage{hyperref}
\usepackage{color}

\begin{document}


\title{Spectral energy-loss bump and $\gamma$-ray pulsar halos}


\author{Kun Fang}
\email[]{fangkun@ihep.ac.cn}
\affiliation{State Key Laboratory of Particle Astrophysics, Institute of High Energy Physics, Chinese Academy of Sciences, Beijing 100049, China}


\date{\today}

\begin{abstract}
The parent electron spectrum of $\gamma$-ray pulsar halos is determined by continuous injection and radiative energy losses, with the nonlinear nature of the latter expected to produce an age-dependent energy-loss bump in the halo spectrum. We demonstrate that the broadband $\gamma$-ray spectrum of the Geminga halo measured by Fermi-LAT and HAWC is consistent with this expectation. Its energy-loss bump appears in the GeV region and is primarily produced by electrons injected at the earliest stage. Therefore, fitting the data allows us to constrain parameters governing early electron injection, such as the pulsar braking index $n$ and initial spin period $P_0$. In addition, we show that another crucial condition for explaining the Geminga halo spectrum is a power-law index of the injection spectrum close to $1$, consistent with the X-ray observations of the Geminga pulsar wind nebula. Currently, no significant energy-loss bump has been detected in the spectrum of the Monogem halo; based on this, we also place preliminary constraints on $n$ and $P_0$ for Monogem. This work demonstrates that pulsar halo spectra can be used not only to constrain electron injection parameters but also to provide independent estimates of certain intrinsic pulsar properties.
\end{abstract}


\maketitle

\section{Introduction}
\label{sec:intro}
When a bow-shock pulsar wind nebula (PWN) traverses the interstellar medium (ISM), it releases relativistic electrons and positrons that diffuse through the ambient turbulent magnetic field. Through inverse Compton scattering (ICS) with background photons, these electrons\footnote{Hereafter, \textit{electrons} refers to both electrons and positrons unless explicitly distinguished.} produce a $\gamma$-ray halo around the pulsar \cite{Abeysekara:2017old}, known as a pulsar halo (or a TeV halo \cite{Linden:2017vvb}). The morphology of the $\gamma$-ray halo is a direct projection of the spatial distribution of the parent electrons, thereby serving as an ideal probe of cosmic-ray transport on scales of $\sim10\ \text{pc}$ \cite{Fang:2022fof,Liu:2022hqf,Lopez-Coto:2022igd,Amato:2024dss}. Stacking analyses suggest that pulsar halos may be common around middle-aged pulsars \cite{Albert:2025gwm}. To date, more than ten confirmed or candidate pulsar halos have been reported \cite{Alfaro:2026tak,Fang:2022qaf,HAWC:2023jsq,LHAASO:2024flo,Wach:2025fqf,DeSarkar:2026eck}\footnote{Most of the reported pulsar halos or candidates have been cataloged in Ref.~\cite{Alfaro:2026tak}. The remaining references are provided as supplements.}.

As pulsar halos are generated by electrons that have escaped from PWNe, they provide a unique probe into the electron injection process from PWNe to the ISM, characterized by the injection rate, energy spectral index, and maximum energy. This is crucial for evaluating the contribution of pulsars to the cosmic-ray positron excess \cite{Abeysekara:2017old,Hooper:2017gtd,Fang:2018qco,Shao-Qiang:2018zla,Zhou:2022jzg,Schroer:2023aoh,Fang:2023xla}.
However, the parent electron spectrum of a pulsar halo can be much more complex than the injection spectrum, as it represents the accumulation of electrons injected at different epochs and subsequently cooled by radiative losses. For the Geminga halo, the prototype of this class, theoretical calculations predict an energy-loss bump in the cumulative electron spectrum at $\sim1\ \text{TeV}$ \cite{Aharonian:1995zz,Yuksel:2008rf}, which corresponds to a $\gamma$-ray spectral bump in the GeV region.
This implies that TeV $\gamma$-ray experiments such as HAWC and LHAASO may miss this critical spectral feature. Moreover, the TeV measurements fall within the high-energy cutoff regime of the halo, where parameter estimation may suffer from significant degeneracies.

Ref.~\cite{DiMauro:2019yvh} searched for the GeV counterpart of the Geminga halo using Fermi-LAT data, reporting a signal in the $10-100\ \text{GeV}$ range after accounting for the pulsar proper motion and adopting a sufficiently large region of interest ($70^\circ\times70^\circ$). Ref.~\cite{Meng:2026tja} independently reproduced the results of Ref.~\cite{DiMauro:2019yvh} and further placed strong constraints on the halo flux below $10\ \text{GeV}$. In this work, we demonstrate that the GeV observations of the Geminga halo are consistent with the spectral features expected from the energy-loss bump in the cumulative electron spectrum. By interpreting the broadband $\gamma$-ray spectrum from $1\ \text{GeV}$ to $100\ \text{TeV}$, we can place valuable constraints not only on the injection spectral parameters but also on certain intrinsic pulsar properties.

In the next section, we describe the origin of the energy-loss bump in the cumulative electron spectrum, its temporal evolution, and how it manifests in the $\gamma$‑ray spectrum of pulsar halos. Section~\ref{sec:geminga} presents constraints on physical parameters from the broadband $\gamma$‑ray spectrum of the Geminga halo, with particular emphasis on the role of the GeV spectral structure in breaking degeneracies among the parameters. In Section~\ref{sec:monogem}, we extend our study to a younger pulsar halo, the Monogem halo, using existing spectral data. Section~\ref{sec:conclu} concludes the paper.

\section{Origin of the spectral energy-loss bump}
\label{sec:bump}

\begin{figure*}[!t]
	\centering 	\includegraphics[width=0.45\textwidth]{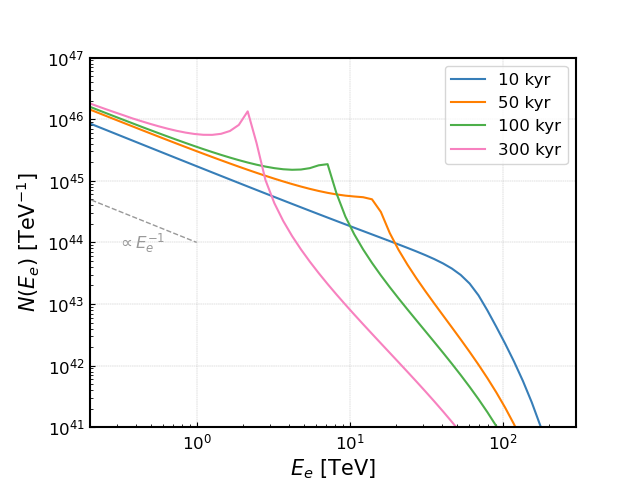}
    \includegraphics[width=0.45\textwidth]{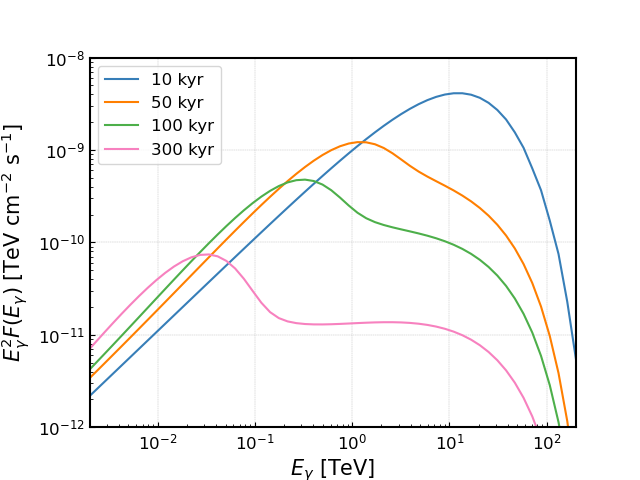}
	\caption{Left: Time evolution of the parent electron spectrum of a pulsar halo, adopting the characteristic age, distance, and spin-down luminosity of Geminga. Other assumed parameters are described in the text. The dashed line indicates the slope of the power-law component in the injection spectrum, indicating that below the energy-loss bump, the shape of the cumulative electron spectrum remains consistent with the injection spectrum. Right: $\gamma$-ray spectra produced through ICS from the corresponding electron spectra shown in the left panel. For clarity, only the CMB is considered as the seed photon field for the ICS process in both subfigures.}
	\label{fig:bump}
\end{figure*}

The energy distribution evolution of the parent electrons of a pulsar halo can be described by the energy-loss equation under continuous injection, written as
\begin{equation}
 \frac{\partial N}{\partial t} = \frac{\partial [b(E_e)N]}{\partial E_e} + Q(E_e, t)\ ,
 \label{eq:loss}
\end{equation}
where $N$ is the energy-differential number of electrons, $E_e$ is the electron kinetic energy, $b(E_e)\equiv|dE_e/dt|$ is the absolute energy-loss rate, and $Q(E_e, t)$ is the source function. We employ \textsc{phect} \cite{Fang:2025eiv}, a computational tool developed for pulsar halos, to solve Eq.~(\ref{eq:loss}) and obtain the pulsar halo emission. We adopt the \texttt{Loss\_A} model in this work, which is suitable for scenarios where only the energy distributions of electrons and $\gamma$ rays are of concern.

The energy-loss rate can be expressed as 
\begin{equation}
 b(E_e)= b_0E_e^2=b_{0,\mathrm{syn}}E_e^2+b_{0,\mathrm{ics}}E_e^2\ .
 \label{eq:b}
\end{equation}
The first term on the far right-hand side of the equation corresponds to the synchrotron loss, where $b_{0,\mathrm{syn}}$ is a constant proportional to the magnetic field energy density, i.e., $b_{0,\mathrm{syn}}\propto B^2$. The second term corresponds to the ICS loss, where $b_{0,\mathrm{ics}}$ is energy-dependent due to the Klein-Nishina (KN) effect \cite{Fang:2020dmi}. The background photon field for ICS losses is identical to that adopted for the subsequent calculation of $\gamma$-ray emission.

The source function $Q(E_e, t)$ can be decomposed into an energy component $q_E$ and a temporal component $q_t$. The energy term, or the electron injection spectrum, takes the form of 
\begin{equation}
 q_E(E_e)\propto E_e^{-p}\ {\rm exp}\left[-\left(\frac{E_e}{E_{e,c}}\right)^2\right]\ ,
 \label{eq:src_e}
\end{equation}
which is suggested by the relativistic shock acceleration theory \cite{Dempsey:2007ng}. The temporal variation of electron injection from the PWN is assumed to follow that of the pulsar spin-down luminosity $L$, as
\begin{equation}
 \begin{aligned}
  q_t(t) \propto L(t) &= L(\tau)\frac{\left(1+\frac{t}{\tau_0}\right)^{-\frac{n+1}{n-1}}}{\left(1+\frac{\tau}{\tau_0}\right)^{-\frac{n+1}{n-1}}} \\
  & = L(\tau)\left(\frac{2\tau_c}{n-1}\right)^{\frac{n+1}{n-1}}(t+\tau_0)^{-\frac{n+1}{n-1}} \ ,
 \end{aligned}
\label{eq:src_t}
\end{equation}
where $t=0$ represents the pulsar birth time, $\tau_c$ is the pulsar characteristic age, $\tau$ is the pulsar true age, $n$ is the pulsar braking index, and $\tau_0$ is the initial spin-down timescale of the pulsar, which can be defined by $2\tau_c/(n-1)-\tau$ \cite{Gaensler:2006ua}. The current spin-down luminosity and characteristic age of Geminga, $L(\tau)=3.2\times10^{34}\ \text{erg}\ \text{s}^{-1}$ and $\tau_c=342\ \text{kyr}$, can be found in the ATNF catalog \cite{Manchester:2004bp}. 

After solving Eq.~(\ref{eq:loss}) for $N$, we obtain the total luminosity $L_\gamma$ of the pulsar halo using the standard isotropic ICS formula \cite{Blumenthal:1970gc}, from which the expected observed flux follows as $F=L_\gamma/(4\pi D^2)$, where $D$ is the distance to the pulsar. For Geminga, trigonometric parallax yields $D=250$~pc \cite{2007Ap&SS.308..225F}.

Using the $L(\tau)$, $\tau_c$, and $D$ values of Geminga, the left panel of Fig.~\ref{fig:bump} shows the temporal evolution of the parent electron spectrum of a pulsar halo. Other assumed parameters include $p=1$, $E_{e,c}=130\ \text{TeV}$, $n=3$ (magnetic dipole radiation), $\tau_0=10\ \text{kyr}$, $B=3\ \mu\text{G}$, and $\eta=5\%$, where $\eta$ is the conversion efficiency from the pulsar spin-down energy to the electron injection energy, primarily determining the normalization of the source function $Q$. For a more straightforward illustration, we consider only the cosmic microwave background (CMB) as the seed photon field for ICS.

For each age $t$, there is a critical energy $E_{e,\text{cr}}$ satisfying 
\begin{equation}
t=\int_{E_{e,\text{cr}}}^\infty dE_e'/b(E_e')\ .
\label{eq:loss_time}
\end{equation}
All electrons injected at time zero have now cooled down to energies below $E_{e,\text{cr}}$. We define $E_e<E_{e,\text{cr}}$ as the injection regime, where the spectrum remains largely unaffected by energy losses and still retains the form of the injected spectrum; $E_e>E_{e,\text{cr}}$ as the energy‑loss regime, where the spectrum has been significantly modified by radiative losses. 

A spectral bump in the electron spectrum is expected around $E_{e,\text{cr}}$ due to the pile-up effect caused by the nonlinear energy losses, i.e., Eq.~(\ref{eq:b}). 
More specifically, all electrons injected at time zero with energies greater than $E_{e,\text{cr}}$ are currently compressed into a narrow energy interval from $\approx E_{e,\text{cr}}/2$ to $E_{e,\text{cr}}$\footnote{Neglecting the energy dependence of $b_0$, one readily obtains $\int_{E_{e,\text{cr}}}^\infty dE_e'/(b_0E_e'^2)=\int_{E_{e,\text{cr}}/2}^{E_{e,\text{cr}}} dE_e'/(b_0E_e'^2)=t$, which means that $t$ can be interpreted either as the timescale for electrons to cool from infinite energy down to $E_{e,\text{cr}}$, or as the timescale for electrons with energy $E_{e,\text{cr}}$ to lose half of their energy.}. Furthermore, according to Eq.~(\ref{eq:src_t}), electrons injected at earlier times are more abundant, resulting in a more pronounced pile-up in the energy spectrum. Consequently, the spectral bump arises from the combined effects of electron energy losses and time-dependent injection.

The left panel of Fig.~\ref{fig:bump} clearly shows the energy-loss bump shifting toward lower energies as the injection time increases. As a simple estimate, neglecting the KN effect of the ICS process, Eq.~(\ref{eq:loss_time}) yields
\begin{equation}
 E_{e,\text{cr}}\approx \frac{1}{b_0t}=\frac{1}{(b_{0,\text{syn}}+b_{0,\text{ics}})t}\ .
 \label{eq:Ecr}
\end{equation}
Given $b_{0,\text{syn}}=2.3\times10^{-14}\ \text{TeV}^{-1}\ \text{s}^{-1}$ and $b_{0,\text{ics}}=2.6\times10^{-14}\ \text{TeV}^{-1}\ \text{s}^{-1}$, $E_{e,\text{cr}}$ is estimated to be $2.2 \ \text{TeV}$ for $t=300\ \text{kyr}$, consistent with the bump position indicated by the pink line in the figure.

The energy-loss bump in the electron spectrum maps onto the $\gamma$-ray spectrum produced through ICS, as shown in the right panel of Fig.~\ref{fig:bump}. For young pulsar halos with electron injection times $\lesssim50\ \text{kyr}$, the energy-loss bump in the $\gamma$-ray spectrum has not yet developed significantly, since $E_{e,\text{cr}}$ has not fallen well below $E_{e,c}$. By the age of Geminga ($\sim$300 kyr), the energy-loss bump has shifted below $100\ \text{GeV}$, achieving clear separation from the high-energy cutoff. If only the spectral measurements above TeV energies are available, the spectra of very young and old halos would exhibit no marked differences, both describable by a power-law component with a high-energy cutoff (the blue and pink lines in the right panel of Fig.~\ref{fig:bump}). However, their broadband spectra are fundamentally different.



\begin{figure*}[!t]
	\centering 	\includegraphics[width=0.95\textwidth]{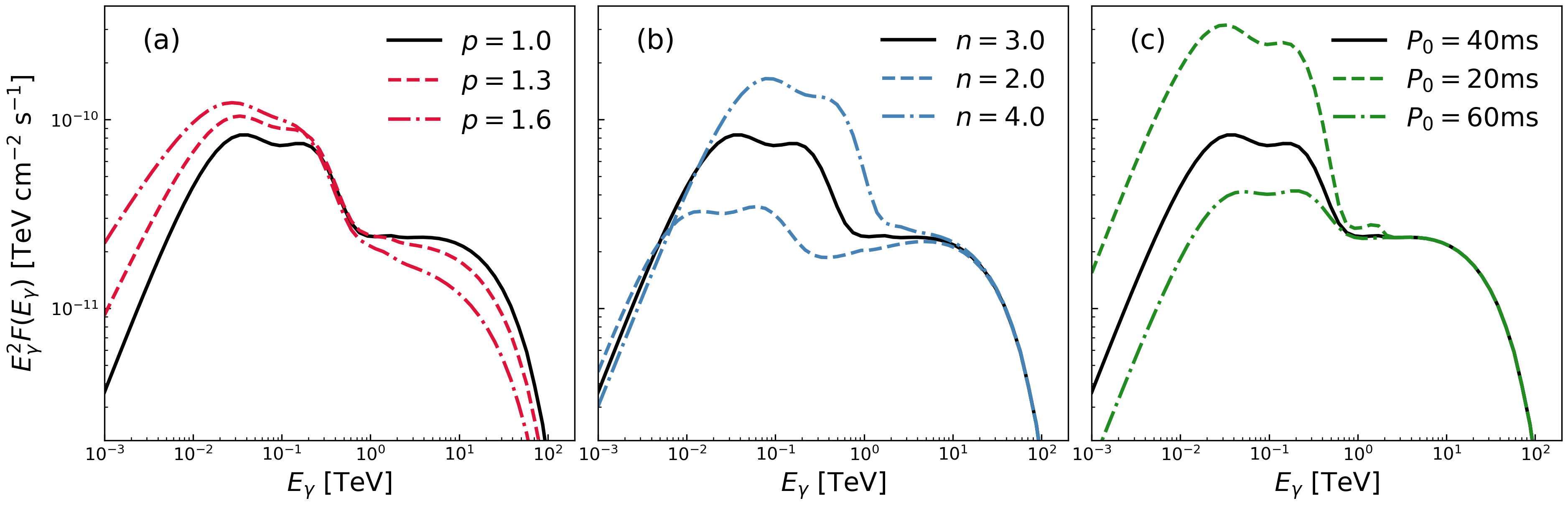}
	\caption{Impact of different parameters on the morphology of the energy-loss bump in the Geminga halo spectrum. (a) Power-law index of the electron injection spectrum. (b) Pulsar braking index. (c) Initial spin period of the pulsar. These parameters affect distinct characteristics of the bump and are therefore not subject to significant degeneracy in the fitting procedure. The energy-loss bump consists of two sub-bumps, with the lower- and higher-energy ones arising from the contributions of the CMB and dust emission, respectively.}
	\label{fig:params}%
\end{figure*}

\section{Broadband \texorpdfstring{$\gamma$}{gamma}-ray spectral fitting of the Geminga halo}
\label{sec:geminga}
In this section, we fit the TeV data from HAWC \cite{HAWC:2024scl} and the GeV data from Fermi-LAT \cite{Meng:2026tja} for the Geminga halo, thereby estimating physical parameters. Before the fitting procedure, we discuss how various parameters affect the halo spectrum to elucidate the basis for the parameter constraints.

\subsection{Impact of parameters on spectral features}
\label{subsec:params}

\begin{itemize}
 \item $\eta$ - With other parameters fixed, the energy conversion efficiency determines the normalization of the injected electron spectrum, thereby affecting the flux level of the $\gamma$-ray spectrum.
 \item $E_{e,c}$ - The cutoff energy of the injection spectrum governs the high-energy roll-off in the $\gamma$-ray spectrum and is therefore constrained by the TeV data.
 \item $p$ - As illustrated in Fig.~\ref{fig:params}a, the power-law index of the injection spectrum affects the width of the energy-loss bump. As discussed in Section~\ref{sec:bump}, the region below the spectral bump corresponds to the injection regime of the cumulative electron spectrum, where the spectral slope is governed by $p$; a smaller $p$ leads to a steeper spectrum in this region. In addition, $p$ also affects the $\gamma$-ray spectral shape above the bump in the TeV range. However, since this region lies near the spectral cutoff and has entered the energy-loss regime of the cumulative electron spectrum, $p$ exhibits degeneracy with parameters such as $E_{e,c}$ and $n$ when only TeV data are considered.
 \item $n$ - As shown in Fig.~\ref{fig:params}b, the braking index of the pulsar affects both the height and the energy position of the energy-loss bump. Since the bump arises from high-energy electrons injected at early times, its height is positively correlated with the initial spin-down luminosity $L_0$. The latter can be expressed as 
 \begin{equation}
  L_0 = L(\tau)\left(\frac{P}{P_0}\right)^{n+1}\ ,
  \label{eq:L0}
 \end{equation}
 where $P_0$ and $P$ are the initial and current spin periods of the pulsar, respectively ($P=0.237\ \text{s}$ for Geminga). With $P_0$ fixed, $L_0$ increases with $n$, and consequently the energy-loss bump becomes more pronounced. Moreover, Eq.~(\ref{eq:Ecr}) indicates that the bump position is affected by the electron injection age. The true age of the pulsar is expressed as
 \begin{equation}
  \tau=\frac{P}{(n-1)\dot{P}}\left[1-\left(\frac{P_0}{P}\right)^{n-1}\right]\ ,
  \label{eq:tau}
 \end{equation}
 where $\dot{P}$ is the current period derivative ($1.1\times10^{-14}$ for Geminga). Since $P_0$ is typically much smaller than $P$, the dependence of $\tau$ on $n$ is dominated by the $1/(n-1)$ factor. Consequently, $\tau$ decreases as $n$ increases, causing the bump to shift toward higher energies.
 \item $P_0$: As shown in Fig.~\ref{fig:params}c, $P_0$ affects the height of the energy-loss bump, which is indicated by Eq.~(\ref{eq:L0}). For an evolved pulsar like Geminga, $P_0$ has little effect on the true pulsar age. This implies that $P_0$ influences only the bump height, with a minor impact on the bump position, and therefore does not degenerate with $n$ during fitting.
\end{itemize}

\subsection{Settings of the fitting procedure}
\label{subsec:settings}
In the following fitting procedure, we treat $\eta$, $E_{e,c}$, $n$, $P_0$, and $p$ as free parameters. We do not use $\tau_0$ as a free parameter in this section. By definition, it is determined by $n$ and $P_0$. With $n$ set free, $P_0$ is a more suitable choice of free parameter because it is independent of $n$. We employ a Bayesian approach for parameter estimation, utilizing the Markov chain Monte Carlo (MCMC) tool \texttt{emcee} \cite{2013PASP..125..306F} to sample the parameter space. Uniform priors are adopted for the five parameters mentioned above, with ranges of $0.1\%-100\%$, $10-1000\ \text{TeV}$, $1.0-5.0$, $5-500\ \text{ms}$, and $1.0-4.0$, respectively. Theoretical considerations and observational evidence both suggest $p\gtrsim1.0$ for the Geminga scenario \cite{Bykov:2017xpo,Posselt:2016lot}; we therefore set a lower bound of $1$ for the prior of $p$.

Currently, no X-ray experiment has detected a counterpart of the Geminga halo \cite{Khokhriakova:2023rqh,Wu:2024ugx,Manconi:2024wlq,Krivonos:2025mpp}, thereby placing constraints on the magnetic field strength near the pulsar. In the fitting presented in this section, we adopt $B=1.4\ \mu\text{G}$, corresponding to the upper limit derived from the degree-wide search by eROSITA \cite{Khokhriakova:2023rqh}. A further weaker magnetic field would not affect our results, as ICS dominates the electron energy losses under this assumption. We adopt the background photon field from Ref.~\cite{Abeysekara:2017old}, assuming it to be composed of the CMB, infrared dust emission, and starlight, with the latter two having temperatures and energy densities of $(20\ \text{K}, 0.3\ \text{eV}\ \text{cm}^{-3})$ and $(5000\ \text{K}, 0.3\ \text{eV}\ \text{cm}^{-3})$, respectively. The contributions of different processes to the energy losses are presented in Appendix~\ref{app:loss}. In the energy range of interest, the contributions from the CMB and dust components to ICS are dominant. As shown in Fig.~\ref{fig:params}, the energy-loss bump in the $\gamma$-ray spectrum actually consists of two sub-bumps, with the lower- and higher-energy ones corresponding to the CMB and the dust emission, respectively.

In the MCMC fitting, we provide the log-likelihood $\log\mathcal{L}=-\chi^2/2$ of the parameters to \texttt{emcee}, where $\chi^2$ is the chi-square statistic between model and data. Both the Fermi-LAT and HAWC datasets include flux upper limits (ULs). We adopt a quadratic penalty term for the ULs \cite{Penek:2021jjm} as 
\begin{equation}
 \Delta\chi^2=\frac{(F_\text{model}-F_\text{UL})^2}{\sigma_\text{UL}^2}\Theta(F_\text{model}-F_\text{UL})\ ,
 \label{eq:quad_penalty}
\end{equation}
where $\Theta$ is the unit step function, and $\sigma_\text{UL}=F_\text{UL}/z_\text{CL}$, with $z_\text{UL}$ the standard normal quantile corresponding to the specified confidence level (CL). HAWC provides ULs with a $90\%$ CL, corresponding to $z_\text{UL}=1.282$; Fermi-LAT provides ULs with a $95\%$ CL, corresponding to $z_\text{UL}=1.645$. The characteristic of Eq.~(\ref{eq:quad_penalty}) is that it allows the model to lie slightly above the UL, while imposing a severe penalty when the model significantly exceeds it\footnote{In comparison, the Naima package \cite{naima} handles ULs using a constant penalty method, that is, a fixed $\log\mathcal{L}$ is assigned regardless of how much the model exceeds the data.}.

\subsection{Fitting results}
\label{subsec:fitting}

\begin{figure}[!t]
	\centering 	\includegraphics[width=0.45\textwidth]{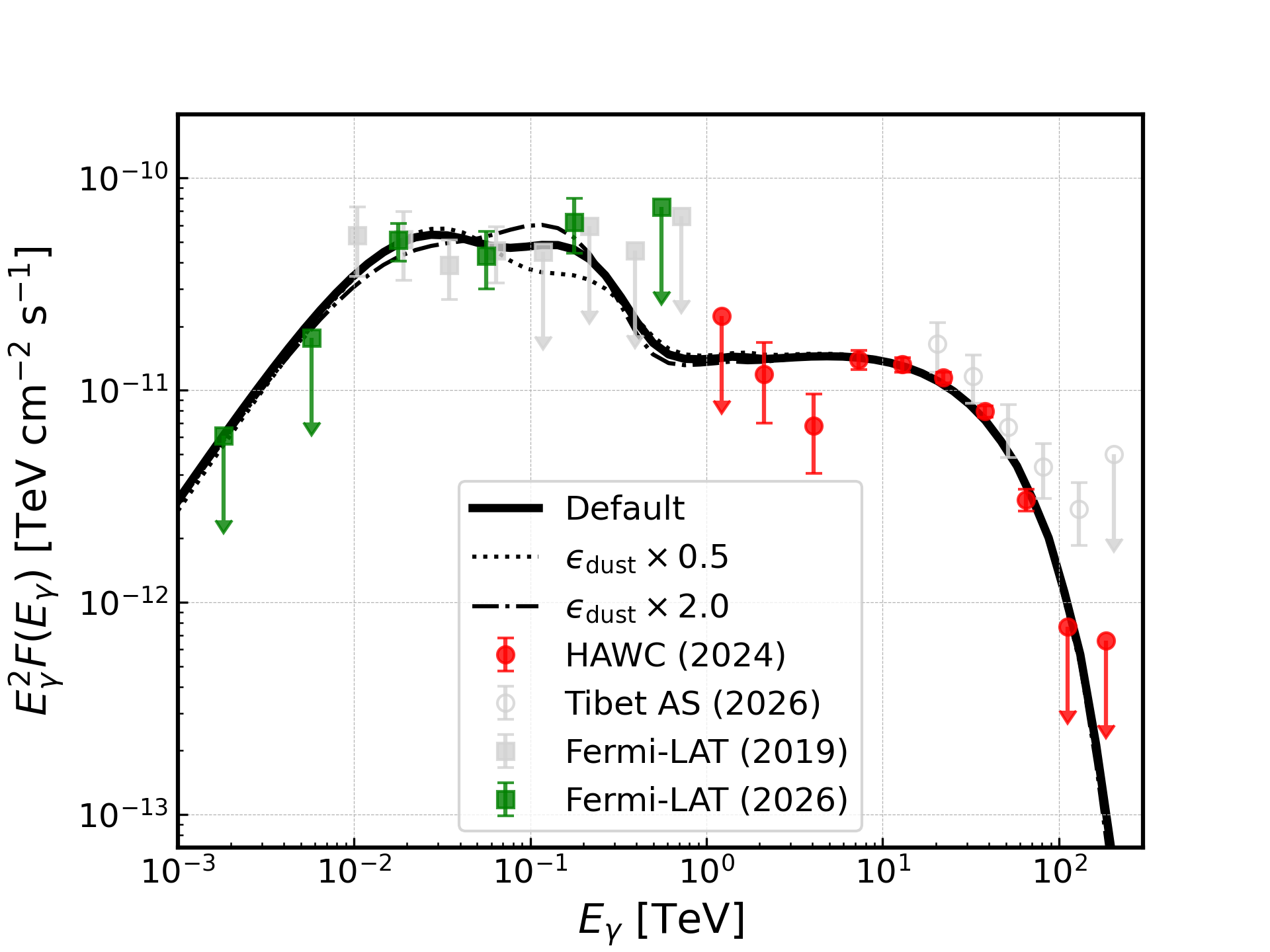}
	\caption{Fit to the broadband $\gamma$-ray spectrum of the Geminga halo. Gray data points are not included in the fit and are shown for comparison only \cite{DiMauro:2019yvh,TibetASg:2026bag}. For the Fermi-LAT data, we adopt the latest analysis results from Ref.~\cite{Meng:2026tja}. For the TeV observations, we use the higher-precision HAWC data \cite{HAWC:2024scl}. The solid curve shows the best fit with default dust emission \cite{Abeysekara:2017old}, while the dotted and dash-dotted curves correspond to dust-emission energy densities halved and doubled, respectively.} 
	\label{fig:gem_bf}%
\end{figure}

\begin{figure*}[!ht]
	\centering 	\includegraphics[width=0.85\textwidth]{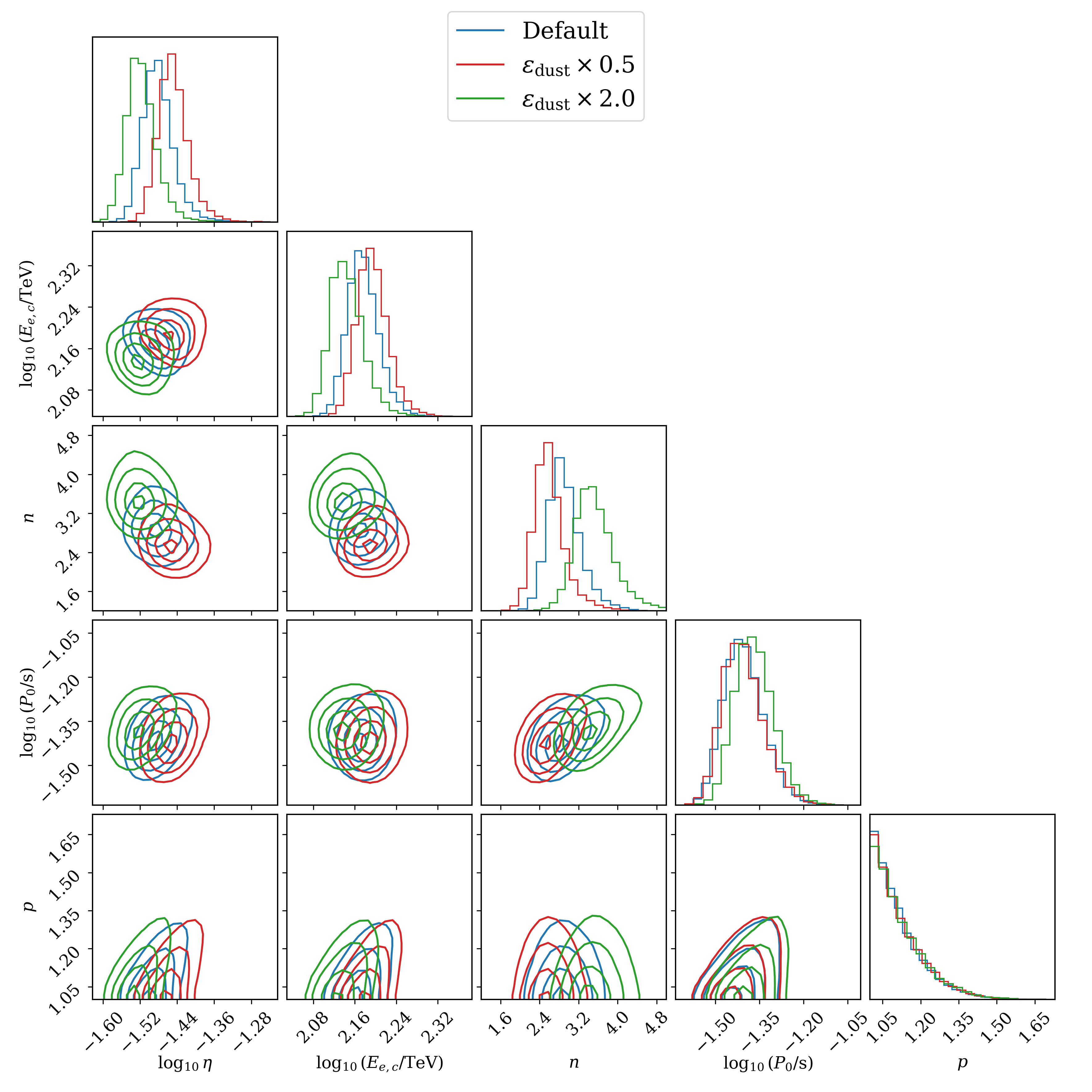}
	\caption{Posterior distributions of physical parameters obtained from fitting the Geminga halo spectrum measured by HAWC and Fermi-LAT. The three sets of distributions correspond to the three models presented in Fig.~\ref{fig:gem_bf}. Numerical summaries of the posterior distributions are presented in Table~\ref{tab:fit}. The priors of the parameters are described in Section~\ref{subsec:settings}.}
	\label{fig:post}%
\end{figure*}

As shown by the solid line in Fig.~\ref{fig:gem_bf}, the best-fit model successfully reproduces the observed broadband spectral features. The best-fit $\chi^2$ is $15$ for $9$ degrees of freedom\footnote{Upper limits contribute to the degrees of freedom only when the $\chi^2$ penalty shown by Eq.~(\ref{eq:quad_penalty}) is activated, as they then constrain the parameter space similarly to data points, providing effective information.}, which falls within an acceptable range ($\text{p-value}\approx0.1$). The parameter estimation results are summarized in Table~\ref{tab:fit}, and the corner plot for the posterior distributions is presented in Fig.~\ref{fig:post}. The energy conversion efficiency $\eta$ is primarily constrained by the HAWC data, because the total energy of the injected electron spectrum is concentrated at the high-energy end for $p<2$ \cite{Malyshev:2009tw}. The fitted $\eta$ is far below $100\%$, a physically reasonable outcome.

Within classical electrodynamics, the maximum particle acceleration capability of a system can be expressed as $E_\text{max}=\varepsilon_\text{acc}eB_\text{acc}R_\text{acc}$, where $B_\text{acc}$ and $R_\text{acc}$ are the magnetic field strength and size of the acceleration zone, respectively, and $\varepsilon_\text{acc}$ is the acceleration efficiency \cite{Aharonian:2002we}. For the Geminga PWN, adopting $B_\text{acc}=20\ \mu\text{G}$ and $R_\text{acc}=0.05\ \text{pc}$ \cite{Posselt:2016lot}, and taking the fitted $E_{e,c}$ as an estimate of $E_\text{max}$, we obtain $\varepsilon_\text{acc}\approx0.15$. This efficiency is comparable to that of the young PWN Crab \cite{LHAASO:2021cbz}, indicating that even an aged bow-shock PWN can maintain high acceleration efficiency. Alternatively, we can evaluate the acceleration efficiency in terms of the maximum potential drop, i.e., $E_\text{max}=\varepsilon_\text{acc}e\sqrt{L/c}$ \cite{Nikonorov:2026bvw}. Substituting $E_\text{max}=140$~TeV and $L=3.2\times10^{34}$~erg~s$^{-1}$ yields $\varepsilon_\text{acc}=0.45$, indicating that the high acceleration efficiency of the Geminga PWN is a robust result. In addition, Ref.~\cite{Nikonorov:2026bvw} argued that the Guitar nebula, a bow-shock PWN older than the Geminga PWN, may reach an even higher efficiency of $\varepsilon_\text{acc}=0.75$.

\begin{table*}[!ht]
\centering
\caption{Fitted and derived parameters for the Geminga halo from Bayesian analyses.}
\begin{ruledtabular}
\begin{tabular}{lcccccc}
\multirow{3}{*}{} & \multicolumn{6}{c}{Energy density of dust emission} \\
\cline{2-7}
 & \multicolumn{2}{c}{Default} & \multicolumn{2}{c}{$\times 0.5$} & \multicolumn{2}{c}{$\times 2$} \\
\cline{2-3}\cline{4-5}\cline{6-7}
 & Best fit & Mean and $1\sigma$ errors & Best fit & Mean and $1\sigma$ errors & Best fit & Mean and $1\sigma$ errors \\
\hline
\multicolumn{7}{l}{Fitted params} \\
\cline{1-1}
$\eta$ [\%] & $3.1$ & $3.3_{-0.2}^{+0.3}$ & $3.4$ & $3.5_{-0.2}^{+0.3}$ & $2.9$ & $3.0_{-0.2}^{+0.2}$ \\
$E_{e,c}$ [TeV] & $144$ & $149_{-9}^{+11}$ & $148$ & $156_{-9}^{+12}$ & $133$ & $139_{-8}^{+11}$ \\
$n$ & $2.80$ & $2.88_{-0.28}^{+0.36}$ & $2.53$ & $2.56_{-0.25}^{+0.33}$ & $3.34$ & $3.49_{-0.34}^{+0.42}$ \\
$P_0$ [ms] & $34$ & $39_{-5}^{+7}$ & $35$ & $39_{-5}^{+7}$ & $38$ & $42_{-6}^{+7}$ \\
$p$ & $1.00$ & $1.08_{-0.06}^{+0.12}$ & $1.00$ & $1.09_{-0.06}^{+0.12}$ & $1.00$ & $1.09_{-0.06}^{+0.12}$ \\
\hline
\multicolumn{7}{l}{Derived params} \\
\cline{1-1}
$L_0$ [$10^{37}$ erg s$^{-1}$] & $5.0$ & $3.6_{-1.6}^{+2.4}$ & $2.7$ & $2.1_{-0.8}^{+1.2}$ & $9.5$ & $7.3_{-3.4}^{+5.6}$ \\
$\tau$ [kyr] & 368 & $351_{-53}^{+54}$ & 423 & $412_{-64}^{+64}$ & 289 & $271_{-38}^{+41}$ \\
\hline
$\chi^2$ & $15.4$ & - & $17.6$ & - & $14.3$ & - \\
\end{tabular}
\end{ruledtabular}
\label{tab:fit}
\end{table*}

Both the braking index and the initial spin period of Geminga are well constrained. The measured $n$ is consistent, within its uncertainties, with the value of $3$ expected for dipole radiation. Currently, $P_0$ estimates are available for only a small number of pulsars, predominantly falling in the range of $10-100\ \text{ms}$ \cite{Migliazzo:2002vd,Suzuki:2021ium}; our fitted value lies within this range. We also infer the initial spin-down luminosity $L_0$ and the true age $\tau$ of Geminga from Eq.~(\ref{eq:L0}) and (\ref{eq:tau}), with $L_0$ an order of magnitude lower than the current $L$ of the Crab pulsar, and $\approx60$ times lower than its inferred $L_0$. Moreover, if $n$ is fixed at $3$, we obtain $P_0=36\ \text{ms}$, yielding an initial spin-down timescale of $\tau_0=8\ \text{kyr}$. This is in close agreement with $\tau_0=9-12\ \text{kyr}$ commonly assumed in pulsar halo studies \cite{Hooper:2017gtd,DiMauro:2019hwn}.

Since the energy-loss bump in the $\gamma$-ray spectrum consists of two sub-bumps, uncertainties in the dust emission also affect the spectral shape. We examine cases where the energy density of the dust emission $\epsilon_\text{dust}$ is reduced or increased by a factor of two relative to the default value, $0.3\ \text{eV}\ \text{cm}^{-3}$; the corresponding fitting results are shown as the dotted and dash-dotted curves in Fig.~\ref{fig:gem_bf}, respectively. Although the best-fit model for the $\epsilon_\text{dust}\times0.5$ case lies remarkably below the data at $\approx200\ \text{GeV}$, its total $\chi^2$ does not show a significant increase over the default result. Future space-borne $\gamma$-ray experiments (e.g., VLAST \cite{Pan:2024adp}), if achieving substantial enhancements in energy resolution and effective area, may be able to independently constrain local dust emission parameters using the Geminga halo spectrum.

As shown in Table~\ref{tab:fit} and Fig.~\ref{fig:post}, reducing or increasing $\epsilon_\text{dust}$ does not significantly affect the parameter estimates, considering the uncertainties. Among these, the $\epsilon_\text{dust}\times0.5$ and $\epsilon_\text{dust}\times2.0$ cases yield relatively different values of $n$, yet both remain consistent with the expectation for pulsar dipole radiation within their uncertainties. In addition, the inferred true age of Geminga differs noticeably across scenarios. The supernova remnant (SNR) associated with Geminga has been proposed as a potentially important local cosmic-ray source (e.g., Ref.~\cite{Qiao:2025zyk}), and its true age is crucial for assessing whether this source can simultaneously account for the cosmic-ray spectral and anisotropy features \cite{Fang:2020cru}.

It can be seen in Fig.~\ref{fig:post} that the posterior distributions of $p$ are concentrated at the lower bound of the prior, $p=1.0$. For such a one-sided distribution, an UL is more statistically meaningful than the mean and $1\sigma$ errors given in Table~\ref{tab:fit}. The $95\%$ credible upper bounds on $p$ for the default, $\epsilon_\text{dust}\times0.5$, and $\epsilon_\text{dust}\times2.0$ cases are $1.31$, $1.31$, and $1.32$, respectively. This indicates that the data require a quite hard power-law component in the electron injection spectrum.

Ref.~\cite{Meng:2026tja} notes that in the Fermi-LAT analysis of the Geminga halo, different Galactic diffuse emission templates yield similar flux ULs below $10\ \text{GeV}$, indicating that the low flux in this energy range is robust. This energy region corresponds to the injection regime of the parent electron spectrum, thus requiring a hard power-law component in the injection spectrum. In addition, the flat spectral feature observed by HAWC in the $1-10\ \text{TeV}$ range also favors a small $p$ \cite{HAWC:2024scl,Fang:2023xla}. This result is consistent with X-ray PWN observations: the two lateral tails of the Geminga X-ray PWN are thought to be produced by electrons injected into the ISM, and the inferred $p$ from their X-ray spectra is around $1.0$ \cite{Posselt:2016lot}.


\subsection{Injection time delay}
\label{subsec:delay}
In the above discussion, we assumed that electron injection begins at $t=0$. In reality, however, the injection may be delayed by $1-10$~kyr. During the early evolution of the PWN-SNR system, the PWN bubble remains relatively isolated from the surrounding SNR ejecta. After several thousand years, the inward-propagating SNR reverse shock crushes the PWN bubble, driving oscillations between compressed and reexpanded states \cite{Blondin:2001bf,2001A&A...380..309V}. At this stage, the relativistic gas within the PWN begins to mix with the ejecta, and electron escape may become significant.

\begin{figure}[!ht]
	\centering 	
    \includegraphics[width=0.48\textwidth]{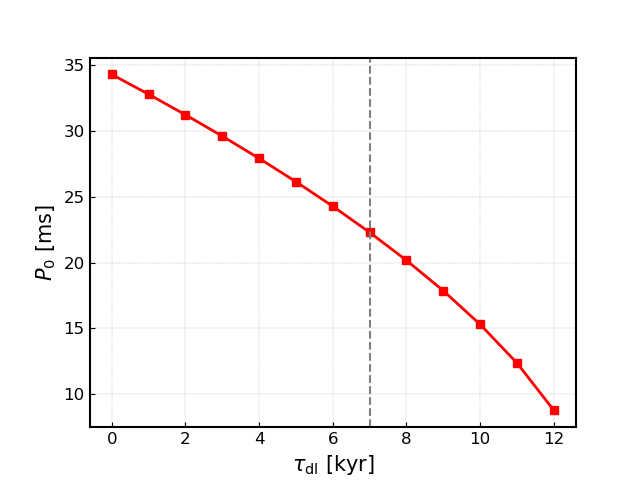}
	\caption{Best-fit $P_0$ as a function of the injection delay time $\tau_\text{dl}$. Both parameters affect the height of the energy-loss bump, leading to strong mutual degeneracy in the spectral fitting of the Geminga halo. The gray dashed line marks the injection delay time adopted by Ref.~\cite{Hinton:2011ad} in their study of electron escape from another famous PWN, the Vela PWN.}
	\label{fig:delay}%
\end{figure}

Since the spectral energy-loss bump originates from electrons injected at early times, the presence of an injection delay alters the expected $\gamma$-ray spectral shape and consequently affects the parameter estimation. We assume different injection delay times $\tau_\text{dl}$ and repeat the fit under the default model described in Section~\ref{subsec:fitting}. The results indicate a strong degeneracy between $\tau_\text{dl}$ and $P_0$, as shown in Fig.~\ref{fig:delay}. This is because the injection delay reduces the expected height of the energy-loss bump, which can be compensated by a smaller $P_0$. The estimates of the other parameters are almost unaffected by $\tau_\text{dl}$.

Figure~\ref{fig:delay} reveals two further interesting results. First, if we assume $\tau_\text{dl}=7\ \text{kyr}$, as adopted in Ref.~\cite{Hinton:2011ad} for the study of electron escape from the Vela PWN, the resulting $P_0$ is $22\ \text{ms}$, very close to the $P_0=19\ \text{ms}$ of the Crab pulsar \cite{Migliazzo:2002vd}. Second, when $\tau_\text{dl}$ reaches $12\ \text{kyr}$, the inferred $P_0$ falls below $10\ \text{ms}$, which is smaller than any currently known $P_0$. To keep the $P_0$ of Geminga within the typical range, $\tau_\text{dl}$ should not significantly exceed $10\ \text{kyr}$, consistent with the PWN-SNR evolution picture \cite{Blondin:2001bf,2001A&A...380..309V}.

\section{The Monogem halo}
\label{sec:monogem}
The Monogem halo is another confirmed pulsar halo, although its broadband $\gamma$-ray spectral characteristics remain uncertain at present, with no significant detection below $\approx5\ \text{TeV}$. Nevertheless, we may still place preliminary parameter constraints.

The characteristic age of the Monogem pulsar is $\tau_c=111\ \text{kyr}$, one-third that of Geminga. Under typical parameters, the energy-loss bump in the expected $\gamma$-ray spectrum of its pulsar halo has shifted below TeV energies, similar to the green curve in the right panel of Fig.~\ref{fig:bump}. In addition, the separation between the energy-loss bump and the high-energy cutoff is less distinct than in the Geminga halo, which results in a softer spectrum above TeV energies. However, Fermi-LAT and HAWC have not observed a clear energy-loss bump or a soft spectral feature above TeV \cite{DiMauro:2019yvh,HAWC:2024scl}.

\begin{figure}[!t]
	\centering 	\includegraphics[width=0.45\textwidth]{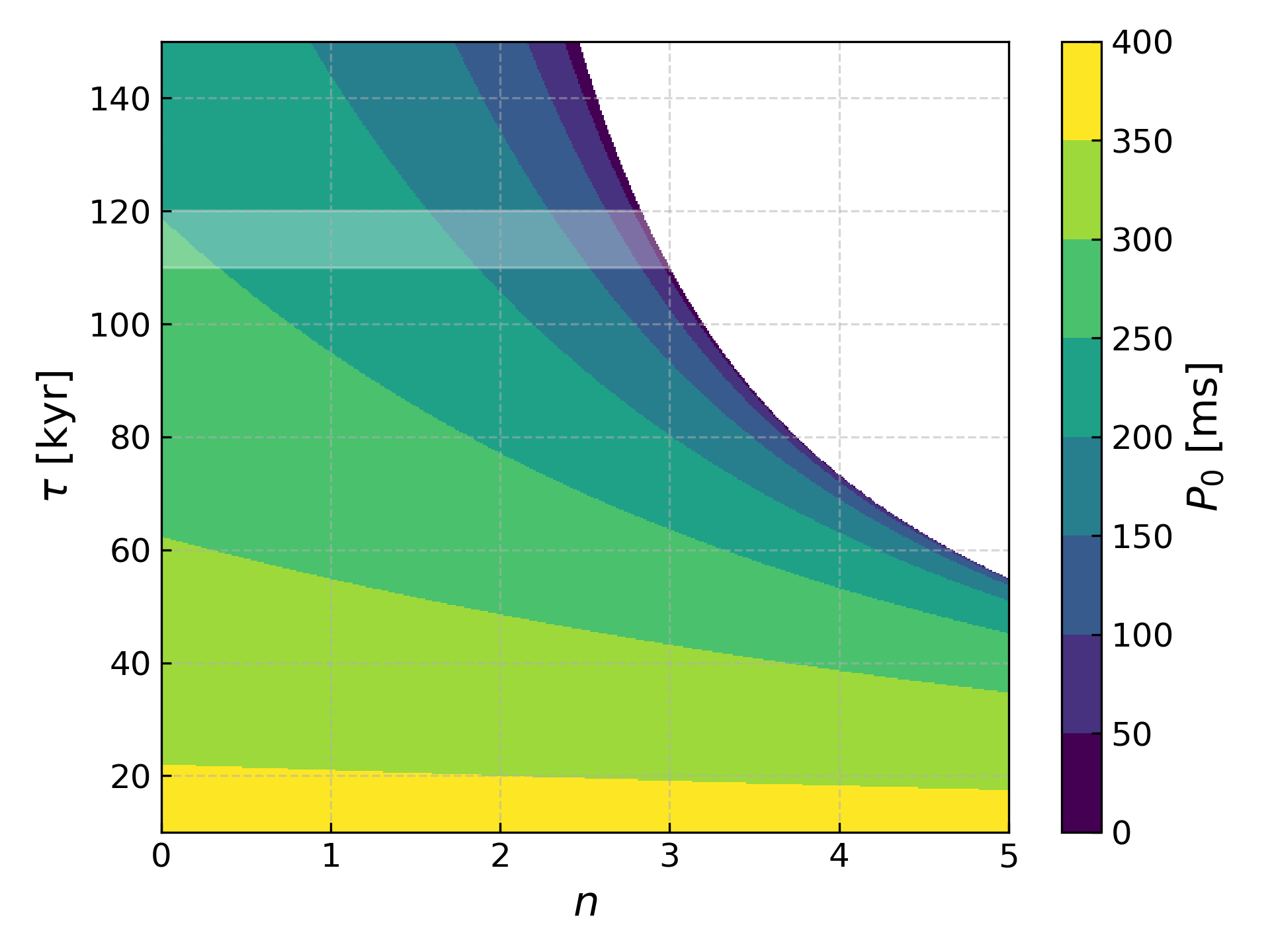}
	\caption{The inferred initial spin period $P_0$ of Monogem as a function of the braking index $n$, for a given true pulsar age $\tau$ and a current spin period of $P=385\ \text{ms}$ \cite{Manchester:2004bp}. The relation follows from Eq.~(\ref{eq:tau}). The white band indicates the latest estimated age range for the Monogem Ring \cite{Knies:2024spk}.} 
	\label{fig:tau_n}%
\end{figure}

Two factors may alleviate the tension between the expectation and the data. First, the energy-loss bump may not shift to sub-TeV energies as expected. According to Eq.~(\ref{eq:Ecr}), a younger pulsar age or a lower energy-loss rate can reduce the shift of the energy-loss bump toward lower energies. However, the latest age estimate for the Monogem Ring, based on eROSITA observations with more careful data analysis and physical modeling, is $110-120\ \text{kyr}$ \cite{Knies:2024spk}, implying that the true age of Monogem should not be significantly smaller than its characteristic age. We therefore adopt a weak magnetic field strength of $1\ \mu\text{G}$ to reduce the electron energy-loss rate.

Second, the energy-loss bump may be less pronounced than expected. As shown in Fig.~\ref{fig:params}, a smaller $n$ or a larger $P_0$ both reduce the bump height. In fact, when the true age $\tau$ is given, a smaller $n$ naturally corresponds to a larger $P_0$, as illustrated in Fig.~\ref{fig:tau_n}. Among pulsars with long-term braking index measurements, Vela has a relatively small and well-constrained braking index of $1.4-1.7$ \cite{1996Natur.381..497L,Espinoza:2016ars}. We adopt $n=1.4$ for Monogem here.

\begin{figure}[!t]
	\centering 	\includegraphics[width=0.45\textwidth]{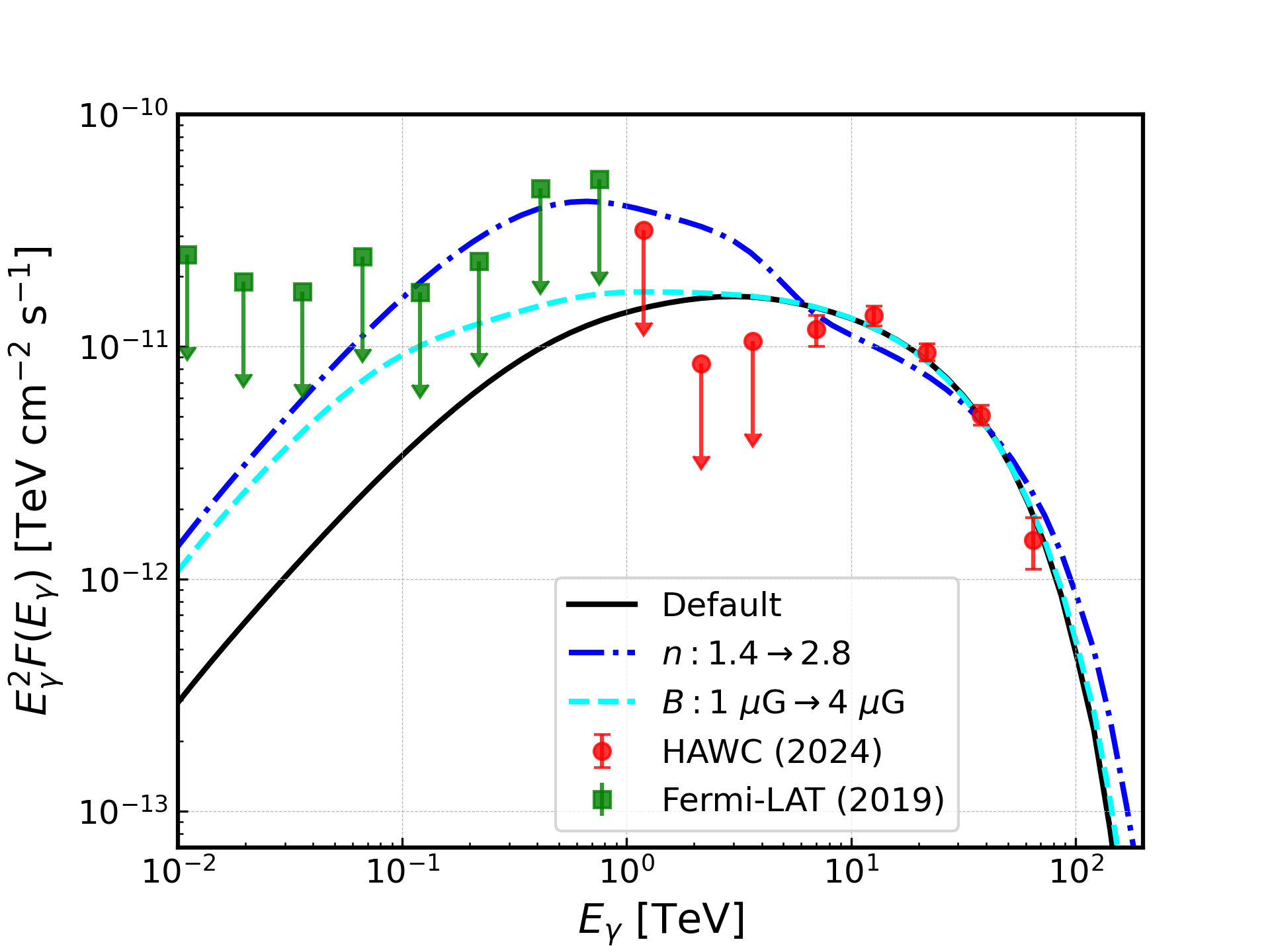}
	\caption{Fit to the broadband $\gamma$-ray spectrum of the Monogem halo. The default curve adopts a small braking index of $n=1.4$ and a weak magnetic field of $B=1\ \mu\text{G}$. These settings suppress the formation of a pronounced energy-loss bump. Also shown are the best-fit curves where $n$ is changed to $2.8$ or $B$ to $4\ \mu\text{G}$ relative to the default model; the former is close to the expectation for pulsar dipole radiation, while the latter corresponds to the UL on the ambient magnetic field strength of Monogem from eROSITA \cite{Khokhriakova:2023rqh}.} 
	\label{fig:mono_bf}%
\end{figure}


\begin{table*}[!ht]
\centering
\caption{Fitted and derived parameters for the Monogem halo from Bayesian analyses.}
\begin{ruledtabular}
\begin{tabular}{lcccccc}
 & \multicolumn{2}{c}{Default} & \multicolumn{2}{c}{$n: 1.4\rightarrow2.8$} & \multicolumn{2}{c}{$B: 1\rightarrow4\ \mu\text{G}$} \\
\cline{2-3}\cline{4-5}\cline{6-7}
 & Best fit & Mean and $1\sigma$ errors & Best fit & Mean and $1\sigma$ errors & Best fit & Mean and $1\sigma$ errors \\
\hline
\multicolumn{7}{l}{Fitted params} \\
\cline{1-1}
$\eta$ [\%] & $2.8$ & $2.8_{-0.2}^{+0.2}$ & $1.8$ & $1.8_{-0.1}^{+0.1}$ & $9.1$ & $9.0_{-0.6}^{+0.7}$ \\
$E_{e,c}$ [TeV] & $108$ & $108_{-7}^{+7}$ & $140$ & $141_{-9}^{+10}$ & $123$ & $123_{-9}^{+10}$ \\
\hline
$\chi^2$ & $8.0$ & - & $37.1$ & - & $9.0$ & - \\
\end{tabular}
\end{ruledtabular}
\label{tab:fit2}
\end{table*}

We perform MCMC fitting to the Monogem halo spectrum from Fermi-LAT and HAWC using the same UL treatment as in Section~\ref{subsec:settings}. We set $\eta$ and $E_{e,c}$ as free parameters and fix $p=1$. Uniform priors are adopted for $\eta$ and $E_{e,c}$, with ranges of $0.1\%-100\%$ and $10-1000\ \text{TeV}$, respectively. The comparison between the best-fit curve and the data is presented in Fig.~\ref{fig:mono_bf}, and the resulting parameter estimates are presented in Table~\ref{tab:fit2}. Although the best-fit model exceeds the HAWC flux ULs in $2-5\ \text{TeV}$, the total $\chi^2$ of $8$ for $5$ degrees of freedom is acceptable ($\text{p-value}\approx0.15$). In Appendix~\ref{app:mono}, we further discuss conditions that could bring the model below those ULs. The energy conversion efficiency required for the Monogem halo is very similar to that for the Geminga halo, both around $3\%$; the cutoff energy of its injection spectrum is slightly lower than that of the Geminga halo.

We have also examined a case with $n$ approaching $3$, with the results shown in Fig.~\ref{fig:mono_bf} and Table~\ref{tab:fit2}. Assuming a true pulsar age of $\tau=115\ \text{kyr}$, the UL on $n$ is $\approx2.9$ according to Fig.~\ref{fig:tau_n}. Adopting $n=2.8$, which is close to the best-fit value under the default model for the Geminga halo, we find that the resulting energy-loss bump is excessively pronounced, leading to significant inconsistency between the model and the HAWC data and raising the $\chi^2$ to $37$. In addition, the flux UL from the eROSITA search for the Monogem halo counterpart corresponds to an UL on the ambient magnetic field strength of $4\ \mu\text{G}$ \cite{Khokhriakova:2023rqh}. Changing $B$ from $1\ \mu\text{G}$ to $4\ \mu\text{G}$, we see from Fig.~\ref{fig:mono_bf} that although the enhanced energy losses broaden the spectrum, no inconsistency with the data arises.

The above analysis indicates that reconciling the model with the data requires $n$ not to approach $3$, which corresponds to $P_0\gtrsim100\ \text{ms}$ according to Fig.~\ref{fig:tau_n}. This implies that the initial spin period of Monogem is larger than that of most pulsars with measured values \cite{Suzuki:2021ium,2025ApJ...987...63A}.

\section{Conclusion}
\label{sec:conclu}
The broadband $\gamma$-ray spectrum of the Geminga halo from $1\ \text{GeV}$ to $100\ \text{TeV}$ can be naturally explained by accounting for continuous injection and radiative energy losses of the parent electrons. The GeV spectral structure measured by Fermi-LAT corresponds to the expected energy-loss bump, which is formed by early-injected electrons that have undergone nonlinear energy losses. For Geminga, whose age is several hundred kyr, the energy-loss bump in its pulsar halo spectrum is already sufficiently separated from the high-energy cutoff. Therefore, by fitting the data, we can constrain the parameters governing early electron injection. The braking index $n$ of the Geminga pulsar is estimated to be $2.3-3.9$, consistent with dipole radiation within its uncertainties. The initial spin period $P_0$ is estimated at $30-50\ \text{ms}$, and it could be smaller if an electron injection delay is considered.

The energy conversion efficiency from pulsar spin-down energy to electron injection energy and the cutoff energy of the injection spectrum are primarily constrained by the HAWC TeV data, yielding values of $\eta\approx3\%$ and $E_{e,c}=130-160\ \text{TeV}$, respectively. In addition, a hard power-law component with an index of $p<1.3$ in the injection spectrum is crucial for explaining the broadband $\gamma$-ray spectral features. This constraint is in good agreement with observations of the Geminga X-ray PWN.

The broadband spectral characteristics of the Monogem halo are currently less well established than those of the Geminga halo; nevertheless, we have placed some preliminary constraints on its parameters. The absence of a pronounced energy-loss bump in its $\gamma$-ray spectrum, combined with the available estimate of the true pulsar age, suggests that the early electron injection rate may be relatively low. This requires that $n$ not approach the dipole-radiation expectation of $3$, corresponding to $P_0\gtrsim100\ \text{ms}$, which exceeds most known pulsar initial spin periods.

This work aims to propose a method for estimating pulsar-related parameters by utilizing the energy-loss bump feature in pulsar halo spectra; at the same time, the quantitative results may be subject to several limitations. Under homogeneous-isotropic diffusion, most injected electrons are concentrated near the pulsar, and the $\gamma$-ray fluxes reported by HAWC and Fermi-LAT approximate the emission from the full spatially integrated electron population. 
In this case, electron spatial propagation need not be considered in spectral studies, as we have done in this work. 
However, given the possible origins of the slow-diffusion phenomenon around pulsars \cite{Evoli:2018aza,Kun:2019sks}, electron diffusion may exhibit two-zone behavior, and the current flux measurements may differ from the true total flux \cite{Fang:2023xla}. More rigorous studies may require iterative interplay between data analysis and modeling of electron propagation. In addition, if the energy conversion efficiency $\eta$ exhibits significant time dependence, parameter estimation could become more complex. Future work may test for time dependence in $\eta$ by studying a sample of pulsar halos with a broad age distribution.

\begin{acknowledgments}
This work is supported by the National Natural Science Foundation of China under Grants No.~12393853 and No.~12105292.
\end{acknowledgments}

\appendix

\section{Energy dependence of the electron radiative energy losses}
\label{app:loss}
\begin{figure}[!ht]
	\centering 	
    \includegraphics[width=0.48\textwidth]{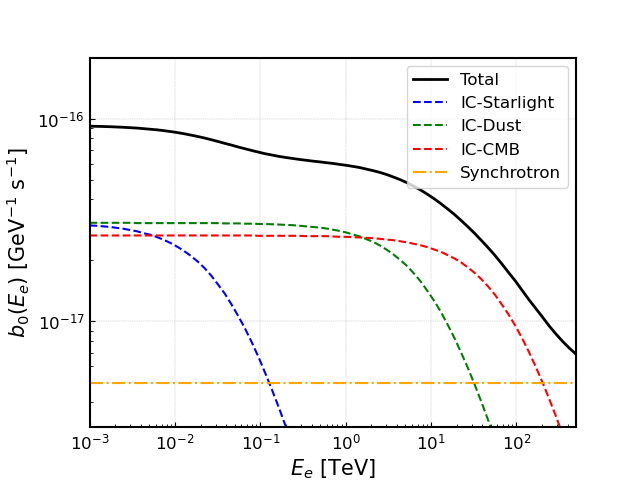}
	\caption{Energy dependence of the energy-loss factor $b_0$. The background photon field and magnetic field settings are described in Section~\ref{subsec:settings}.} 
	\label{fig:loss}%
\end{figure}

Figure~\ref{fig:loss} shows the energy dependence of the energy-loss factor $b_0$, reflecting the relative contributions of different processes to the electron energy loss. The parent electron energy considered in this work is mainly between $0.1-100\ \text{TeV}$, where the CMB and dust components of ICS dominate the energy loss. The starlight component of ICS is suppressed by the KN effect above $\sim10$~GeV, so its contribution to either the electron energy loss or the $\gamma$-ray emission is insignificant. In addition, X-ray observations impose a low magnetic field constraint on the environment of Geminga, making the synchrotron loss subdominant below $\sim100\ \text{TeV}$.

\section{Potential large curvature of the Monogram halo spectrum}
\label{app:mono}
To reconcile the calculated Monogem halo spectrum with the flux ULs in the $2–5\ \text{TeV}$ range, significant spectral curvature is required. As shown in the right panel of Fig.~\ref{fig:bump}, this would require a young source age; otherwise, early-injected electrons would broaden the spectrum through energy losses. However, as discussed above, the age of the Monogem SNR has been independently estimated, so one must instead assume a significant delay in electron injection. We therefore treat the injection delay time $\tau_\text{dl}$ as a free parameter to fit the data, based on the default model presented in Section~\ref{sec:monogem}. In addition, to minimize electron energy losses, we consider the most optimistic scenario where only the CMB is retained as the seed photon field for ICS.

\begin{figure}[!ht]
	\centering 	
    \includegraphics[width=0.48\textwidth]{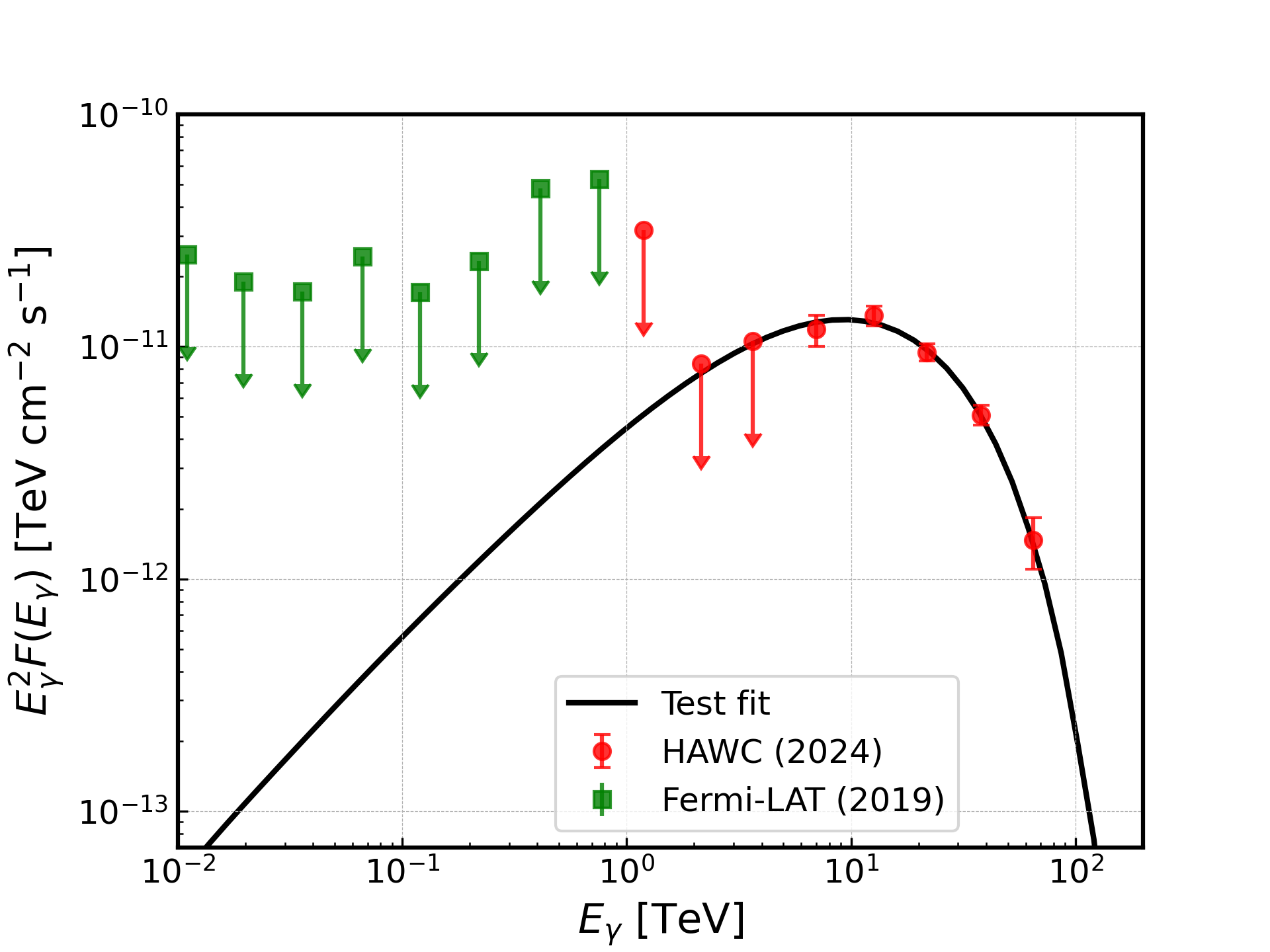}
	\caption{Test fit result for the Monogem halo subject to the HAWC flux ULs in $2-5\ \text{TeV}$. The ICS seed photon field is restricted to the CMB. The electron injection delay $\tau_\text{dl}$ is set as a free parameter. The curve corresponds to the best fit with $\tau_\text{dl}=90\ \text{kyr}$, which is unphysically large.}
	\label{fig:mono_opt}%
\end{figure}

The test fit result is presented in Fig.~\ref{fig:mono_opt}. Although the best-fit spectrum satisfies the HAWC ULs, the fit yields $\tau_\text{dl}=90\pm12\ \text{kyr}$, which is an unphysically large injection delay time. The current flux ULs in the $2–5\ \text{TeV}$ range may require cross-checking with future measurements from other experiments.

\bibliography{references}

\begin{thebibliography}{60}%
\makeatletter
\providecommand \@ifxundefined [1]{%
 \@ifx{#1\undefined}
}%
\providecommand \@ifnum [1]{%
 \ifnum #1\expandafter \@firstoftwo
 \else \expandafter \@secondoftwo
 \fi
}%
\providecommand \@ifx [1]{%
 \ifx #1\expandafter \@firstoftwo
 \else \expandafter \@secondoftwo
 \fi
}%
\providecommand \natexlab [1]{#1}%
\providecommand \enquote  [1]{``#1''}%
\providecommand \bibnamefont  [1]{#1}%
\providecommand \bibfnamefont [1]{#1}%
\providecommand \citenamefont [1]{#1}%
\providecommand \href@noop [0]{\@secondoftwo}%
\providecommand \href [0]{\begingroup \@sanitize@url \@href}%
\providecommand \@href[1]{\@@startlink{#1}\@@href}%
\providecommand \@@href[1]{\endgroup#1\@@endlink}%
\providecommand \@sanitize@url [0]{\catcode `\\12\catcode `\$12\catcode
  `\&12\catcode `\#12\catcode `\^12\catcode `\_12\catcode `\%12\relax}%
\providecommand \@@startlink[1]{}%
\providecommand \@@endlink[0]{}%
\providecommand \url  [0]{\begingroup\@sanitize@url \@url }%
\providecommand \@url [1]{\endgroup\@href {#1}{\urlprefix }}%
\providecommand \urlprefix  [0]{URL }%
\providecommand \Eprint [0]{\href }%
\providecommand \doibase [0]{https://doi.org/}%
\providecommand \selectlanguage [0]{\@gobble}%
\providecommand \bibinfo  [0]{\@secondoftwo}%
\providecommand \bibfield  [0]{\@secondoftwo}%
\providecommand \translation [1]{[#1]}%
\providecommand \BibitemOpen [0]{}%
\providecommand \bibitemStop [0]{}%
\providecommand \bibitemNoStop [0]{.\EOS\space}%
\providecommand \EOS [0]{\spacefactor3000\relax}%
\providecommand \BibitemShut  [1]{\csname bibitem#1\endcsname}%
\let\auto@bib@innerbib\@empty
\bibitem [{\citenamefont {Abeysekara}\ \emph {et~al.}(2017)\citenamefont
  {Abeysekara} \emph {et~al.}}]{Abeysekara:2017old}%
  \BibitemOpen
  \bibfield  {author} {\bibinfo {author} {\bibfnamefont {A.}~\bibnamefont
  {Abeysekara}} \emph {et~al.} (\bibinfo {collaboration} {HAWC}),\ }\href
  {https://doi.org/10.1126/science.aan4880} {\bibfield  {journal} {\bibinfo
  {journal} {Science}\ }\textbf {\bibinfo {volume} {358}},\ \bibinfo {pages}
  {911} (\bibinfo {year} {2017})},\ \Eprint {https://arxiv.org/abs/1711.06223}
  {arXiv:1711.06223 [astro-ph.HE]} \BibitemShut {NoStop}%
\bibitem [{\citenamefont {Linden}\ \emph {et~al.}(2017)\citenamefont {Linden},
  \citenamefont {Auchettl}, \citenamefont {Bramante}, \citenamefont {Cholis},
  \citenamefont {Fang}, \citenamefont {Hooper}, \citenamefont {Karwal},\ and\
  \citenamefont {Li}}]{Linden:2017vvb}%
  \BibitemOpen
  \bibfield  {author} {\bibinfo {author} {\bibfnamefont {T.}~\bibnamefont
  {Linden}}, \bibinfo {author} {\bibfnamefont {K.}~\bibnamefont {Auchettl}},
  \bibinfo {author} {\bibfnamefont {J.}~\bibnamefont {Bramante}}, \bibinfo
  {author} {\bibfnamefont {I.}~\bibnamefont {Cholis}}, \bibinfo {author}
  {\bibfnamefont {K.}~\bibnamefont {Fang}}, \bibinfo {author} {\bibfnamefont
  {D.}~\bibnamefont {Hooper}}, \bibinfo {author} {\bibfnamefont
  {T.}~\bibnamefont {Karwal}},\ and\ \bibinfo {author} {\bibfnamefont {S.~W.}\
  \bibnamefont {Li}},\ }\href {https://doi.org/10.1103/PhysRevD.96.103016}
  {\bibfield  {journal} {\bibinfo  {journal} {Phys. Rev. D}\ }\textbf {\bibinfo
  {volume} {96}},\ \bibinfo {pages} {103016} (\bibinfo {year} {2017})},\
  \Eprint {https://arxiv.org/abs/1703.09704} {arXiv:1703.09704 [astro-ph.HE]}
  \BibitemShut {NoStop}%
\bibitem [{\citenamefont {Fang}(2022)}]{Fang:2022fof}%
  \BibitemOpen
  \bibfield  {author} {\bibinfo {author} {\bibfnamefont {K.}~\bibnamefont
  {Fang}},\ }\href {https://doi.org/10.3389/fspas.2022.1022100} {\bibfield
  {journal} {\bibinfo  {journal} {Front. Astron. Space Sci.}\ }\textbf
  {\bibinfo {volume} {9}},\ \bibinfo {pages} {1022100} (\bibinfo {year}
  {2022})},\ \Eprint {https://arxiv.org/abs/2209.13294} {arXiv:2209.13294
  [astro-ph.HE]} \BibitemShut {NoStop}%
\bibitem [{\citenamefont {Liu}(2022)}]{Liu:2022hqf}%
  \BibitemOpen
  \bibfield  {author} {\bibinfo {author} {\bibfnamefont {R.-Y.}\ \bibnamefont
  {Liu}},\ }\href {https://doi.org/10.1142/S0217751X22300113} {\bibfield
  {journal} {\bibinfo  {journal} {Int. J. Mod. Phys. A}\ }\textbf {\bibinfo
  {volume} {37}},\ \bibinfo {pages} {2230011} (\bibinfo {year} {2022})},\
  \Eprint {https://arxiv.org/abs/2207.04011} {arXiv:2207.04011 [astro-ph.HE]}
  \BibitemShut {NoStop}%
\bibitem [{\citenamefont {L\'opez-Coto}\ \emph {et~al.}(2022)\citenamefont
  {L\'opez-Coto}, \citenamefont {de~O\~na Wilhelmi}, \citenamefont {Aharonian},
  \citenamefont {Amato},\ and\ \citenamefont {Hinton}}]{Lopez-Coto:2022igd}%
  \BibitemOpen
  \bibfield  {author} {\bibinfo {author} {\bibfnamefont {R.}~\bibnamefont
  {L\'opez-Coto}}, \bibinfo {author} {\bibfnamefont {E.}~\bibnamefont {de~O\~na
  Wilhelmi}}, \bibinfo {author} {\bibfnamefont {F.}~\bibnamefont {Aharonian}},
  \bibinfo {author} {\bibfnamefont {E.}~\bibnamefont {Amato}},\ and\ \bibinfo
  {author} {\bibfnamefont {J.}~\bibnamefont {Hinton}},\ }\href
  {https://doi.org/10.1038/s41550-021-01580-0} {\bibfield  {journal} {\bibinfo
  {journal} {Nature Astron.}\ }\textbf {\bibinfo {volume} {6}},\ \bibinfo
  {pages} {199} (\bibinfo {year} {2022})},\ \Eprint
  {https://arxiv.org/abs/2202.06899} {arXiv:2202.06899 [astro-ph.HE]}
  \BibitemShut {NoStop}%
\bibitem [{\citenamefont {Amato}\ and\ \citenamefont
  {Recchia}(2024)}]{Amato:2024dss}%
  \BibitemOpen
  \bibfield  {author} {\bibinfo {author} {\bibfnamefont {E.}~\bibnamefont
  {Amato}}\ and\ \bibinfo {author} {\bibfnamefont {S.}~\bibnamefont
  {Recchia}},\ }\href {https://doi.org/10.1007/s40766-024-00059-8} {\bibfield
  {journal} {\bibinfo  {journal} {Riv. Nuovo Cim.}\ }\textbf {\bibinfo {volume}
  {47}},\ \bibinfo {pages} {399} (\bibinfo {year} {2024})},\ \Eprint
  {https://arxiv.org/abs/2409.00659} {arXiv:2409.00659 [astro-ph.HE]}
  \BibitemShut {NoStop}%
\bibitem [{\citenamefont {Albert}\ \emph {et~al.}(2025)\citenamefont {Albert}
  \emph {et~al.}}]{Albert:2025gwm}%
  \BibitemOpen
  \bibfield  {author} {\bibinfo {author} {\bibfnamefont {A.}~\bibnamefont
  {Albert}} \emph {et~al.},\ }\href
  {https://doi.org/10.1103/PhysRevLett.134.171005} {\bibfield  {journal}
  {\bibinfo  {journal} {Phys. Rev. Lett.}\ }\textbf {\bibinfo {volume} {134}},\
  \bibinfo {pages} {171005} (\bibinfo {year} {2025})},\ \Eprint
  {https://arxiv.org/abs/2505.00175} {arXiv:2505.00175 [astro-ph.HE]}
  \BibitemShut {NoStop}%
\bibitem [{\citenamefont {Alfaro}\ \emph {et~al.}(2026)\citenamefont {Alfaro}
  \emph {et~al.}}]{Alfaro:2026tak}%
  \BibitemOpen
  \bibfield  {author} {\bibinfo {author} {\bibfnamefont {R.}~\bibnamefont
  {Alfaro}} \emph {et~al.},\ }\bibfield  {journal} {\bibinfo  {journal} {arXiv
  e-prints}\ }\href {https://doi.org/10.48550/arXiv.2602.00263}
  {10.48550/arXiv.2602.00263} (\bibinfo {year} {2026}),\ \Eprint
  {https://arxiv.org/abs/2602.00263} {arXiv:2602.00263 [astro-ph.HE]}
  \BibitemShut {NoStop}%
\bibitem [{\citenamefont {Fang}\ \emph {et~al.}(2022)\citenamefont {Fang},
  \citenamefont {Xi}, \citenamefont {Bao}, \citenamefont {Bi},\ and\
  \citenamefont {Chen}}]{Fang:2022qaf}%
  \BibitemOpen
  \bibfield  {author} {\bibinfo {author} {\bibfnamefont {K.}~\bibnamefont
  {Fang}}, \bibinfo {author} {\bibfnamefont {S.-Q.}\ \bibnamefont {Xi}},
  \bibinfo {author} {\bibfnamefont {L.-Z.}\ \bibnamefont {Bao}}, \bibinfo
  {author} {\bibfnamefont {X.-J.}\ \bibnamefont {Bi}},\ and\ \bibinfo {author}
  {\bibfnamefont {E.-S.}\ \bibnamefont {Chen}},\ }\href
  {https://doi.org/10.1103/PhysRevD.106.123017} {\bibfield  {journal} {\bibinfo
   {journal} {Phys. Rev. D}\ }\textbf {\bibinfo {volume} {106}},\ \bibinfo
  {pages} {123017} (\bibinfo {year} {2022})},\ \Eprint
  {https://arxiv.org/abs/2207.13533} {arXiv:2207.13533 [astro-ph.HE]}
  \BibitemShut {NoStop}%
\bibitem [{\citenamefont {Albert}\ \emph {et~al.}(2023)\citenamefont {Albert}
  \emph {et~al.}}]{HAWC:2023jsq}%
  \BibitemOpen
  \bibfield  {author} {\bibinfo {author} {\bibfnamefont {A.}~\bibnamefont
  {Albert}} \emph {et~al.} (\bibinfo {collaboration} {HAWC}),\ }\href
  {https://doi.org/10.3847/2041-8213/acb5ee} {\bibfield  {journal} {\bibinfo
  {journal} {Astrophys. J. Lett.}\ }\textbf {\bibinfo {volume} {944}},\
  \bibinfo {pages} {L29} (\bibinfo {year} {2023})},\ \Eprint
  {https://arxiv.org/abs/2301.04646} {arXiv:2301.04646 [astro-ph.HE]}
  \BibitemShut {NoStop}%
\bibitem [{\citenamefont {Cao}\ \emph {et~al.}(2025)\citenamefont {Cao} \emph
  {et~al.}}]{LHAASO:2024flo}%
  \BibitemOpen
  \bibfield  {author} {\bibinfo {author} {\bibfnamefont {Z.}~\bibnamefont
  {Cao}} \emph {et~al.} (\bibinfo {collaboration} {LHAASO}),\ }\href
  {https://doi.org/10.1007/s11433-024-2508-5} {\bibfield  {journal} {\bibinfo
  {journal} {Sci. China Phys. Mech. Astron.}\ }\textbf {\bibinfo {volume}
  {68}},\ \bibinfo {pages} {279504} (\bibinfo {year} {2025})},\ \Eprint
  {https://arxiv.org/abs/2410.04425} {arXiv:2410.04425 [astro-ph.HE]}
  \BibitemShut {NoStop}%
\bibitem [{\citenamefont {Wach}\ and\ \citenamefont
  {Mitchell}(2025)}]{Wach:2025fqf}%
  \BibitemOpen
  \bibfield  {author} {\bibinfo {author} {\bibfnamefont {T.}~\bibnamefont
  {Wach}}\ and\ \bibinfo {author} {\bibfnamefont {A.~M.~W.}\ \bibnamefont
  {Mitchell}},\ }\href {https://doi.org/10.22323/1.501.0874} {\bibfield
  {journal} {\bibinfo  {journal} {PoS}\ }\textbf {\bibinfo {volume}
  {ICRC2025}},\ \bibinfo {pages} {874} (\bibinfo {year} {2025})},\ \Eprint
  {https://arxiv.org/abs/2510.02802} {arXiv:2510.02802 [astro-ph.HE]}
  \BibitemShut {NoStop}%
\bibitem [{\citenamefont {De~Sarkar}(2026)}]{DeSarkar:2026eck}%
  \BibitemOpen
  \bibfield  {author} {\bibinfo {author} {\bibfnamefont {A.}~\bibnamefont
  {De~Sarkar}},\ }\href {https://doi.org/10.1051/0004-6361/202557618}
  {\bibfield  {journal} {\bibinfo  {journal} {Astron. Astrophys.}\ }\textbf
  {\bibinfo {volume} {706}},\ \bibinfo {pages} {A378} (\bibinfo {year}
  {2026})},\ \Eprint {https://arxiv.org/abs/2601.21689} {arXiv:2601.21689
  [astro-ph.HE]} \BibitemShut {NoStop}%
\bibitem [{\citenamefont {Hooper}\ \emph {et~al.}(2017)\citenamefont {Hooper},
  \citenamefont {Cholis}, \citenamefont {Linden},\ and\ \citenamefont
  {Fang}}]{Hooper:2017gtd}%
  \BibitemOpen
  \bibfield  {author} {\bibinfo {author} {\bibfnamefont {D.}~\bibnamefont
  {Hooper}}, \bibinfo {author} {\bibfnamefont {I.}~\bibnamefont {Cholis}},
  \bibinfo {author} {\bibfnamefont {T.}~\bibnamefont {Linden}},\ and\ \bibinfo
  {author} {\bibfnamefont {K.}~\bibnamefont {Fang}},\ }\href
  {https://doi.org/10.1103/PhysRevD.96.103013} {\bibfield  {journal} {\bibinfo
  {journal} {Phys. Rev. D}\ }\textbf {\bibinfo {volume} {96}},\ \bibinfo
  {pages} {103013} (\bibinfo {year} {2017})},\ \Eprint
  {https://arxiv.org/abs/1702.08436} {arXiv:1702.08436 [astro-ph.HE]}
  \BibitemShut {NoStop}%
\bibitem [{\citenamefont {Fang}\ \emph {et~al.}(2018)\citenamefont {Fang},
  \citenamefont {Bi}, \citenamefont {Yin},\ and\ \citenamefont
  {Yuan}}]{Fang:2018qco}%
  \BibitemOpen
  \bibfield  {author} {\bibinfo {author} {\bibfnamefont {K.}~\bibnamefont
  {Fang}}, \bibinfo {author} {\bibfnamefont {X.-J.}\ \bibnamefont {Bi}},
  \bibinfo {author} {\bibfnamefont {P.-F.}\ \bibnamefont {Yin}},\ and\ \bibinfo
  {author} {\bibfnamefont {Q.}~\bibnamefont {Yuan}},\ }\href
  {https://doi.org/10.3847/1538-4357/aad092} {\bibfield  {journal} {\bibinfo
  {journal} {Astrophys. J.}\ }\textbf {\bibinfo {volume} {863}},\ \bibinfo
  {pages} {30} (\bibinfo {year} {2018})},\ \Eprint
  {https://arxiv.org/abs/1803.02640} {arXiv:1803.02640 [astro-ph.HE]}
  \BibitemShut {NoStop}%
\bibitem [{\citenamefont {Xi}\ \emph {et~al.}(2019)\citenamefont {Xi},
  \citenamefont {Liu}, \citenamefont {Huang}, \citenamefont {Fang},\ and\
  \citenamefont {Wang}}]{Shao-Qiang:2018zla}%
  \BibitemOpen
  \bibfield  {author} {\bibinfo {author} {\bibfnamefont {S.-Q.}\ \bibnamefont
  {Xi}}, \bibinfo {author} {\bibfnamefont {R.-Y.}\ \bibnamefont {Liu}},
  \bibinfo {author} {\bibfnamefont {Z.-Q.}\ \bibnamefont {Huang}}, \bibinfo
  {author} {\bibfnamefont {K.}~\bibnamefont {Fang}},\ and\ \bibinfo {author}
  {\bibfnamefont {X.-Y.}\ \bibnamefont {Wang}},\ }\href
  {https://doi.org/10.3847/1538-4357/ab20c9} {\bibfield  {journal} {\bibinfo
  {journal} {Astrophys. J.}\ }\textbf {\bibinfo {volume} {878}},\ \bibinfo
  {pages} {104} (\bibinfo {year} {2019})},\ \Eprint
  {https://arxiv.org/abs/1810.10928} {arXiv:1810.10928 [astro-ph.HE]}
  \BibitemShut {NoStop}%
\bibitem [{\citenamefont {Zhou}\ \emph {et~al.}(2022)\citenamefont {Zhou},
  \citenamefont {Yu}, \citenamefont {Yuan},\ and\ \citenamefont
  {Zhang}}]{Zhou:2022jzg}%
  \BibitemOpen
  \bibfield  {author} {\bibinfo {author} {\bibfnamefont {G.-Y.}\ \bibnamefont
  {Zhou}}, \bibinfo {author} {\bibfnamefont {Z.-H.}\ \bibnamefont {Yu}},
  \bibinfo {author} {\bibfnamefont {Q.}~\bibnamefont {Yuan}},\ and\ \bibinfo
  {author} {\bibfnamefont {H.-H.}\ \bibnamefont {Zhang}},\ }\href
  {https://doi.org/10.1088/1572-9494/ac7cd6} {\bibfield  {journal} {\bibinfo
  {journal} {Commun. Theor. Phys.}\ }\textbf {\bibinfo {volume} {74}},\
  \bibinfo {pages} {105403} (\bibinfo {year} {2022})},\ \Eprint
  {https://arxiv.org/abs/2205.07038} {arXiv:2205.07038 [astro-ph.HE]}
  \BibitemShut {NoStop}%
\bibitem [{\citenamefont {Schroer}\ \emph {et~al.}(2023)\citenamefont
  {Schroer}, \citenamefont {Evoli},\ and\ \citenamefont
  {Blasi}}]{Schroer:2023aoh}%
  \BibitemOpen
  \bibfield  {author} {\bibinfo {author} {\bibfnamefont {B.}~\bibnamefont
  {Schroer}}, \bibinfo {author} {\bibfnamefont {C.}~\bibnamefont {Evoli}},\
  and\ \bibinfo {author} {\bibfnamefont {P.}~\bibnamefont {Blasi}},\ }\href
  {https://doi.org/10.1103/PhysRevD.107.123020} {\bibfield  {journal} {\bibinfo
   {journal} {Phys. Rev. D}\ }\textbf {\bibinfo {volume} {107}},\ \bibinfo
  {pages} {123020} (\bibinfo {year} {2023})},\ \Eprint
  {https://arxiv.org/abs/2305.08019} {arXiv:2305.08019 [astro-ph.HE]}
  \BibitemShut {NoStop}%
\bibitem [{\citenamefont {Fang}(2024)}]{Fang:2023xla}%
  \BibitemOpen
  \bibfield  {author} {\bibinfo {author} {\bibfnamefont {K.}~\bibnamefont
  {Fang}},\ }\href {https://doi.org/10.1103/PhysRevD.109.043041} {\bibfield
  {journal} {\bibinfo  {journal} {Phys. Rev. D}\ }\textbf {\bibinfo {volume}
  {109}},\ \bibinfo {pages} {043041} (\bibinfo {year} {2024})},\ \Eprint
  {https://arxiv.org/abs/2310.16594} {arXiv:2310.16594 [astro-ph.HE]}
  \BibitemShut {NoStop}%
\bibitem [{\citenamefont {Aharonian}\ \emph {et~al.}(1995)\citenamefont
  {Aharonian}, \citenamefont {Atoyan},\ and\ \citenamefont
  {Volk}}]{Aharonian:1995zz}%
  \BibitemOpen
  \bibfield  {author} {\bibinfo {author} {\bibfnamefont {F.~A.}\ \bibnamefont
  {Aharonian}}, \bibinfo {author} {\bibfnamefont {A.~M.}\ \bibnamefont
  {Atoyan}},\ and\ \bibinfo {author} {\bibfnamefont {H.~J.}\ \bibnamefont
  {Volk}},\ }\href@noop {} {\bibfield  {journal} {\bibinfo  {journal} {Astron.
  Astrophys.}\ }\textbf {\bibinfo {volume} {294}},\ \bibinfo {pages} {L41}
  (\bibinfo {year} {1995})}\BibitemShut {NoStop}%
\bibitem [{\citenamefont {Yuksel}\ \emph {et~al.}(2009)\citenamefont {Yuksel},
  \citenamefont {Kistler},\ and\ \citenamefont {Stanev}}]{Yuksel:2008rf}%
  \BibitemOpen
  \bibfield  {author} {\bibinfo {author} {\bibfnamefont {H.}~\bibnamefont
  {Yuksel}}, \bibinfo {author} {\bibfnamefont {M.~D.}\ \bibnamefont
  {Kistler}},\ and\ \bibinfo {author} {\bibfnamefont {T.}~\bibnamefont
  {Stanev}},\ }\href {https://doi.org/10.1103/PhysRevLett.103.051101}
  {\bibfield  {journal} {\bibinfo  {journal} {Phys. Rev. Lett.}\ }\textbf
  {\bibinfo {volume} {103}},\ \bibinfo {pages} {051101} (\bibinfo {year}
  {2009})},\ \Eprint {https://arxiv.org/abs/0810.2784} {arXiv:0810.2784
  [astro-ph]} \BibitemShut {NoStop}%
\bibitem [{\citenamefont {Di~Mauro}\ \emph {et~al.}(2019)\citenamefont
  {Di~Mauro}, \citenamefont {Manconi},\ and\ \citenamefont
  {Donato}}]{DiMauro:2019yvh}%
  \BibitemOpen
  \bibfield  {author} {\bibinfo {author} {\bibfnamefont {M.}~\bibnamefont
  {Di~Mauro}}, \bibinfo {author} {\bibfnamefont {S.}~\bibnamefont {Manconi}},\
  and\ \bibinfo {author} {\bibfnamefont {F.}~\bibnamefont {Donato}},\ }\href
  {https://doi.org/10.1103/PhysRevD.104.089903} {\bibfield  {journal} {\bibinfo
   {journal} {Phys. Rev. D}\ }\textbf {\bibinfo {volume} {100}},\ \bibinfo
  {pages} {123015} (\bibinfo {year} {2019})},\ \bibinfo {note} {[Erratum:
  Phys.Rev.D 104, 089903 (2021)]},\ \Eprint {https://arxiv.org/abs/1903.05647}
  {arXiv:1903.05647 [astro-ph.HE]} \BibitemShut {NoStop}%
\bibitem [{\citenamefont {Meng}\ \emph {et~al.}(2026)\citenamefont {Meng},
  \citenamefont {Liang}, \citenamefont {Zhu}, \citenamefont {Sun},\ and\
  \citenamefont {Liang}}]{Meng:2026tja}%
  \BibitemOpen
  \bibfield  {author} {\bibinfo {author} {\bibfnamefont {Y.}~\bibnamefont
  {Meng}}, \bibinfo {author} {\bibfnamefont {Y.-F.}\ \bibnamefont {Liang}},
  \bibinfo {author} {\bibfnamefont {B.-Y.}\ \bibnamefont {Zhu}}, \bibinfo
  {author} {\bibfnamefont {X.-N.}\ \bibnamefont {Sun}},\ and\ \bibinfo {author}
  {\bibfnamefont {E.-W.}\ \bibnamefont {Liang}},\ }\href
  {https://doi.org/10.1103/qlyv-qh6r} {\bibfield  {journal} {\bibinfo
  {journal} {Phys. Rev. D}\ }\textbf {\bibinfo {volume} {113}},\ \bibinfo
  {pages} {063027} (\bibinfo {year} {2026})},\ \Eprint
  {https://arxiv.org/abs/2602.20882} {arXiv:2602.20882 [astro-ph.HE]}
  \BibitemShut {NoStop}%
\bibitem [{\citenamefont {Fang}(2026)}]{Fang:2025eiv}%
  \BibitemOpen
  \bibfield  {author} {\bibinfo {author} {\bibfnamefont {K.}~\bibnamefont
  {Fang}},\ }\href {https://doi.org/10.1103/lxmg-75p8} {\bibfield  {journal}
  {\bibinfo  {journal} {Phys. Rev. D}\ }\textbf {\bibinfo {volume} {113}},\
  \bibinfo {pages} {083018} (\bibinfo {year} {2026})},\ \Eprint
  {https://arxiv.org/abs/2508.13667} {arXiv:2508.13667 [astro-ph.HE]}
  \BibitemShut {NoStop}%
\bibitem [{\citenamefont {Fang}\ \emph {et~al.}(2021)\citenamefont {Fang},
  \citenamefont {Bi}, \citenamefont {Lin},\ and\ \citenamefont
  {Yuan}}]{Fang:2020dmi}%
  \BibitemOpen
  \bibfield  {author} {\bibinfo {author} {\bibfnamefont {K.}~\bibnamefont
  {Fang}}, \bibinfo {author} {\bibfnamefont {X.-J.}\ \bibnamefont {Bi}},
  \bibinfo {author} {\bibfnamefont {S.-J.}\ \bibnamefont {Lin}},\ and\ \bibinfo
  {author} {\bibfnamefont {Q.}~\bibnamefont {Yuan}},\ }\href
  {https://doi.org/10.1088/0256-307X/38/3/039801} {\bibfield  {journal}
  {\bibinfo  {journal} {Chin. Phys. Lett.}\ }\textbf {\bibinfo {volume} {38}},\
  \bibinfo {pages} {039801} (\bibinfo {year} {2021})},\ \Eprint
  {https://arxiv.org/abs/2007.15601} {arXiv:2007.15601 [astro-ph.HE]}
  \BibitemShut {NoStop}%
\bibitem [{\citenamefont {Dempsey}\ and\ \citenamefont
  {Duffy}(2007)}]{Dempsey:2007ng}%
  \BibitemOpen
  \bibfield  {author} {\bibinfo {author} {\bibfnamefont {P.}~\bibnamefont
  {Dempsey}}\ and\ \bibinfo {author} {\bibfnamefont {P.}~\bibnamefont
  {Duffy}},\ }\href {https://doi.org/10.1111/j.1365-2966.2007.11800.x}
  {\bibfield  {journal} {\bibinfo  {journal} {Mon. Not. Roy. Astron. Soc.}\
  }\textbf {\bibinfo {volume} {378}},\ \bibinfo {pages} {625} (\bibinfo {year}
  {2007})},\ \Eprint {https://arxiv.org/abs/0704.0168} {arXiv:0704.0168
  [astro-ph]} \BibitemShut {NoStop}%
\bibitem [{\citenamefont {Gaensler}\ and\ \citenamefont
  {Slane}(2006)}]{Gaensler:2006ua}%
  \BibitemOpen
  \bibfield  {author} {\bibinfo {author} {\bibfnamefont {B.~M.}\ \bibnamefont
  {Gaensler}}\ and\ \bibinfo {author} {\bibfnamefont {P.~O.}\ \bibnamefont
  {Slane}},\ }\href {https://doi.org/10.1146/annurev.astro.44.051905.092528}
  {\bibfield  {journal} {\bibinfo  {journal} {Ann. Rev. Astron. Astrophys.}\
  }\textbf {\bibinfo {volume} {44}},\ \bibinfo {pages} {17} (\bibinfo {year}
  {2006})},\ \Eprint {https://arxiv.org/abs/astro-ph/0601081}
  {arXiv:astro-ph/0601081} \BibitemShut {NoStop}%
\bibitem [{\citenamefont {Manchester}\ \emph {et~al.}(2005)\citenamefont
  {Manchester}, \citenamefont {Hobbs}, \citenamefont {Teoh},\ and\
  \citenamefont {Hobbs}}]{Manchester:2004bp}%
  \BibitemOpen
  \bibfield  {author} {\bibinfo {author} {\bibfnamefont {R.~N.}\ \bibnamefont
  {Manchester}}, \bibinfo {author} {\bibfnamefont {G.~B.}\ \bibnamefont
  {Hobbs}}, \bibinfo {author} {\bibfnamefont {A.}~\bibnamefont {Teoh}},\ and\
  \bibinfo {author} {\bibfnamefont {M.}~\bibnamefont {Hobbs}},\ }\href
  {https://doi.org/10.1086/428488} {\bibfield  {journal} {\bibinfo  {journal}
  {Astron. J.}\ }\textbf {\bibinfo {volume} {129}},\ \bibinfo {pages} {1993}
  (\bibinfo {year} {2005})},\ \Eprint {https://arxiv.org/abs/astro-ph/0412641}
  {arXiv:astro-ph/0412641} \BibitemShut {NoStop}%
\bibitem [{\citenamefont {Blumenthal}\ and\ \citenamefont
  {Gould}(1970)}]{Blumenthal:1970gc}%
  \BibitemOpen
  \bibfield  {author} {\bibinfo {author} {\bibfnamefont {G.}~\bibnamefont
  {Blumenthal}}\ and\ \bibinfo {author} {\bibfnamefont {R.}~\bibnamefont
  {Gould}},\ }\href {https://doi.org/10.1103/RevModPhys.42.237} {\bibfield
  {journal} {\bibinfo  {journal} {Rev. Mod. Phys.}\ }\textbf {\bibinfo {volume}
  {42}},\ \bibinfo {pages} {237} (\bibinfo {year} {1970})}\BibitemShut
  {NoStop}%
\bibitem [{\citenamefont {{Faherty}}\ \emph {et~al.}(2007)\citenamefont
  {{Faherty}}, \citenamefont {{Walter}},\ and\ \citenamefont
  {{Anderson}}}]{2007Ap&SS.308..225F}%
  \BibitemOpen
  \bibfield  {author} {\bibinfo {author} {\bibfnamefont {J.}~\bibnamefont
  {{Faherty}}}, \bibinfo {author} {\bibfnamefont {F.~M.}\ \bibnamefont
  {{Walter}}},\ and\ \bibinfo {author} {\bibfnamefont {J.}~\bibnamefont
  {{Anderson}}},\ }\href {https://doi.org/10.1007/s10509-007-9368-0} {\bibfield
   {journal} {\bibinfo  {journal} {Astrophys. Space Sci.}\ }\textbf {\bibinfo
  {volume} {308}},\ \bibinfo {pages} {225} (\bibinfo {year}
  {2007})}\BibitemShut {NoStop}%
\bibitem [{\citenamefont {Albert}\ \emph {et~al.}(2024)\citenamefont {Albert}
  \emph {et~al.}}]{HAWC:2024scl}%
  \BibitemOpen
  \bibfield  {author} {\bibinfo {author} {\bibfnamefont {A.}~\bibnamefont
  {Albert}} \emph {et~al.} (\bibinfo {collaboration} {HAWC}),\ }\href
  {https://doi.org/10.3847/1538-4357/ad738e} {\bibfield  {journal} {\bibinfo
  {journal} {Astrophys. J.}\ }\textbf {\bibinfo {volume} {974}},\ \bibinfo
  {pages} {246} (\bibinfo {year} {2024})}\BibitemShut {NoStop}%
\bibitem [{\citenamefont {{Foreman-Mackey}}\ \emph {et~al.}(2013)\citenamefont
  {{Foreman-Mackey}}, \citenamefont {{Hogg}}, \citenamefont {{Lang}},\ and\
  \citenamefont {{Goodman}}}]{2013PASP..125..306F}%
  \BibitemOpen
  \bibfield  {author} {\bibinfo {author} {\bibfnamefont {D.}~\bibnamefont
  {{Foreman-Mackey}}}, \bibinfo {author} {\bibfnamefont {D.~W.}\ \bibnamefont
  {{Hogg}}}, \bibinfo {author} {\bibfnamefont {D.}~\bibnamefont {{Lang}}},\
  and\ \bibinfo {author} {\bibfnamefont {J.}~\bibnamefont {{Goodman}}},\ }\href
  {https://doi.org/10.1086/670067} {\bibfield  {journal} {\bibinfo  {journal}
  {Publications of the Astronomical Society of the Pacific}\ }\textbf {\bibinfo
  {volume} {125}},\ \bibinfo {pages} {306} (\bibinfo {year} {2013})},\ \Eprint
  {https://arxiv.org/abs/1202.3665} {arXiv:1202.3665 [astro-ph.IM]}
  \BibitemShut {NoStop}%
\bibitem [{\citenamefont {Bykov}\ \emph {et~al.}(2017)\citenamefont {Bykov},
  \citenamefont {Amato}, \citenamefont {Petrov}, \citenamefont
  {Krassilchtchikov},\ and\ \citenamefont {Levenfish}}]{Bykov:2017xpo}%
  \BibitemOpen
  \bibfield  {author} {\bibinfo {author} {\bibfnamefont {A.~M.}\ \bibnamefont
  {Bykov}}, \bibinfo {author} {\bibfnamefont {E.}~\bibnamefont {Amato}},
  \bibinfo {author} {\bibfnamefont {A.~E.}\ \bibnamefont {Petrov}}, \bibinfo
  {author} {\bibfnamefont {A.~M.}\ \bibnamefont {Krassilchtchikov}},\ and\
  \bibinfo {author} {\bibfnamefont {K.~P.}\ \bibnamefont {Levenfish}},\ }\href
  {https://doi.org/10.1007/s11214-017-0371-7} {\bibfield  {journal} {\bibinfo
  {journal} {Space Sci. Rev.}\ }\textbf {\bibinfo {volume} {207}},\ \bibinfo
  {pages} {235} (\bibinfo {year} {2017})},\ \Eprint
  {https://arxiv.org/abs/1705.00950} {arXiv:1705.00950 [astro-ph.HE]}
  \BibitemShut {NoStop}%
\bibitem [{\citenamefont {Posselt}\ \emph {et~al.}(2017)\citenamefont
  {Posselt}, \citenamefont {Pavlov}, \citenamefont {Slane}, \citenamefont
  {Romani}, \citenamefont {Bucciantini}, \citenamefont {Bykov}, \citenamefont
  {Kargaltsev}, \citenamefont {Weisskopf},\ and\ \citenamefont
  {Ng}}]{Posselt:2016lot}%
  \BibitemOpen
  \bibfield  {author} {\bibinfo {author} {\bibfnamefont {B.}~\bibnamefont
  {Posselt}}, \bibinfo {author} {\bibfnamefont {G.}~\bibnamefont {Pavlov}},
  \bibinfo {author} {\bibfnamefont {P.}~\bibnamefont {Slane}}, \bibinfo
  {author} {\bibfnamefont {R.}~\bibnamefont {Romani}}, \bibinfo {author}
  {\bibfnamefont {N.}~\bibnamefont {Bucciantini}}, \bibinfo {author}
  {\bibfnamefont {A.}~\bibnamefont {Bykov}}, \bibinfo {author} {\bibfnamefont
  {O.}~\bibnamefont {Kargaltsev}}, \bibinfo {author} {\bibfnamefont
  {M.}~\bibnamefont {Weisskopf}},\ and\ \bibinfo {author} {\bibfnamefont
  {C.~Y.}\ \bibnamefont {Ng}},\ }\href
  {https://doi.org/10.3847/1538-4357/835/1/66} {\bibfield  {journal} {\bibinfo
  {journal} {Astrophys. J.}\ }\textbf {\bibinfo {volume} {835}},\ \bibinfo
  {pages} {66} (\bibinfo {year} {2017})},\ \Eprint
  {https://arxiv.org/abs/1611.03496} {arXiv:1611.03496 [astro-ph.HE]}
  \BibitemShut {NoStop}%
\bibitem [{\citenamefont {Khokhriakova}\ \emph {et~al.}(2024)\citenamefont
  {Khokhriakova}, \citenamefont {Becker}, \citenamefont {Ponti}, \citenamefont
  {Sasaki}, \citenamefont {Li},\ and\ \citenamefont
  {Liu}}]{Khokhriakova:2023rqh}%
  \BibitemOpen
  \bibfield  {author} {\bibinfo {author} {\bibfnamefont {A.}~\bibnamefont
  {Khokhriakova}}, \bibinfo {author} {\bibfnamefont {W.}~\bibnamefont
  {Becker}}, \bibinfo {author} {\bibfnamefont {G.}~\bibnamefont {Ponti}},
  \bibinfo {author} {\bibfnamefont {M.}~\bibnamefont {Sasaki}}, \bibinfo
  {author} {\bibfnamefont {B.}~\bibnamefont {Li}},\ and\ \bibinfo {author}
  {\bibfnamefont {R.~Y.}\ \bibnamefont {Liu}},\ }\href
  {https://doi.org/10.1051/0004-6361/202347311} {\bibfield  {journal} {\bibinfo
   {journal} {Astron. Astrophys.}\ }\textbf {\bibinfo {volume} {683}},\
  \bibinfo {pages} {A180} (\bibinfo {year} {2024})},\ \Eprint
  {https://arxiv.org/abs/2310.10454} {arXiv:2310.10454 [astro-ph.HE]}
  \BibitemShut {NoStop}%
\bibitem [{\citenamefont {Wu}\ \emph {et~al.}(2024)\citenamefont {Wu},
  \citenamefont {Li}, \citenamefont {Liang}, \citenamefont {Ge},\ and\
  \citenamefont {Liu}}]{Wu:2024ugx}%
  \BibitemOpen
  \bibfield  {author} {\bibinfo {author} {\bibfnamefont {Q.-Z.}\ \bibnamefont
  {Wu}}, \bibinfo {author} {\bibfnamefont {C.-M.}\ \bibnamefont {Li}}, \bibinfo
  {author} {\bibfnamefont {X.-H.}\ \bibnamefont {Liang}}, \bibinfo {author}
  {\bibfnamefont {C.}~\bibnamefont {Ge}},\ and\ \bibinfo {author}
  {\bibfnamefont {R.-Y.}\ \bibnamefont {Liu}},\ }\href
  {https://doi.org/10.3847/1538-4357/ad43e1} {\bibfield  {journal} {\bibinfo
  {journal} {Astrophys. J.}\ }\textbf {\bibinfo {volume} {969}},\ \bibinfo
  {pages} {9} (\bibinfo {year} {2024})},\ \Eprint
  {https://arxiv.org/abs/2401.17982} {arXiv:2401.17982 [astro-ph.HE]}
  \BibitemShut {NoStop}%
\bibitem [{\citenamefont {Manconi}\ \emph {et~al.}(2024)\citenamefont
  {Manconi}, \citenamefont {Woo}, \citenamefont {Shang}, \citenamefont
  {Krivonos}, \citenamefont {Tang}, \citenamefont {Di~Mauro}, \citenamefont
  {Donato}, \citenamefont {Mori},\ and\ \citenamefont
  {Hailey}}]{Manconi:2024wlq}%
  \BibitemOpen
  \bibfield  {author} {\bibinfo {author} {\bibfnamefont {S.}~\bibnamefont
  {Manconi}}, \bibinfo {author} {\bibfnamefont {J.}~\bibnamefont {Woo}},
  \bibinfo {author} {\bibfnamefont {R.-Y.}\ \bibnamefont {Shang}}, \bibinfo
  {author} {\bibfnamefont {R.}~\bibnamefont {Krivonos}}, \bibinfo {author}
  {\bibfnamefont {C.}~\bibnamefont {Tang}}, \bibinfo {author} {\bibfnamefont
  {M.}~\bibnamefont {Di~Mauro}}, \bibinfo {author} {\bibfnamefont
  {F.}~\bibnamefont {Donato}}, \bibinfo {author} {\bibfnamefont
  {K.}~\bibnamefont {Mori}},\ and\ \bibinfo {author} {\bibfnamefont {C.~J.}\
  \bibnamefont {Hailey}},\ }\href {https://doi.org/10.1051/0004-6361/202450242}
  {\bibfield  {journal} {\bibinfo  {journal} {Astron. Astrophys.}\ }\textbf
  {\bibinfo {volume} {689}},\ \bibinfo {pages} {A326} (\bibinfo {year}
  {2024})},\ \Eprint {https://arxiv.org/abs/2403.10902} {arXiv:2403.10902
  [astro-ph.HE]} \BibitemShut {NoStop}%
\bibitem [{\citenamefont {Krivonos}\ \emph {et~al.}(2026)\citenamefont
  {Krivonos} \emph {et~al.}}]{Krivonos:2025mpp}%
  \BibitemOpen
  \bibfield  {author} {\bibinfo {author} {\bibfnamefont {R.}~\bibnamefont
  {Krivonos}} \emph {et~al.},\ }\href
  {https://doi.org/10.1051/0004-6361/202555650} {\bibfield  {journal} {\bibinfo
   {journal} {Astron. Astrophys.}\ }\textbf {\bibinfo {volume} {705}},\
  \bibinfo {pages} {A107} (\bibinfo {year} {2026})},\ \Eprint
  {https://arxiv.org/abs/2512.13846} {arXiv:2512.13846 [astro-ph.HE]}
  \BibitemShut {NoStop}%
\bibitem [{\citenamefont {Penek}(2021)}]{Penek:2021jjm}%
  \BibitemOpen
  \bibfield  {author} {\bibinfo {author} {\bibfnamefont {{\"O}.~V.~L.}\
  \bibnamefont {Penek}, \bibfnamefont {Livia}},\ }\emph {\bibinfo {title}
  {{Measurement of pp and CNO cycle solar neutrinos with Borexino}}},\ \href
  {https://doi.org/10.18154/RWTH-2021-08588} {Ph.D. thesis},\ \bibinfo
  {school} {Rheinisch-Westf{\"a}lischen Technischen Hochschule (RWTH) Aachen}
  (\bibinfo {year} {2021})\BibitemShut {NoStop}%
\bibitem [{\citenamefont {{Zabalza}}(2015)}]{naima}%
  \BibitemOpen
  \bibfield  {author} {\bibinfo {author} {\bibfnamefont {V.}~\bibnamefont
  {{Zabalza}}},\ }\href@noop {} {\bibfield  {journal} {\bibinfo  {journal}
  {Proc.~of International Cosmic Ray Conference 2015}\ ,\ \bibinfo {pages}
  {922}} (\bibinfo {year} {2015})},\ \Eprint {https://arxiv.org/abs/1509.03319}
  {1509.03319} \BibitemShut {NoStop}%
\bibitem [{\citenamefont {Malyshev}\ \emph {et~al.}(2009)\citenamefont
  {Malyshev}, \citenamefont {Cholis},\ and\ \citenamefont
  {Gelfand}}]{Malyshev:2009tw}%
  \BibitemOpen
  \bibfield  {author} {\bibinfo {author} {\bibfnamefont {D.}~\bibnamefont
  {Malyshev}}, \bibinfo {author} {\bibfnamefont {I.}~\bibnamefont {Cholis}},\
  and\ \bibinfo {author} {\bibfnamefont {J.}~\bibnamefont {Gelfand}},\ }\href
  {https://doi.org/10.1103/PhysRevD.80.063005} {\bibfield  {journal} {\bibinfo
  {journal} {Phys. Rev. D}\ }\textbf {\bibinfo {volume} {80}},\ \bibinfo
  {pages} {063005} (\bibinfo {year} {2009})},\ \Eprint
  {https://arxiv.org/abs/0903.1310} {arXiv:0903.1310 [astro-ph.HE]}
  \BibitemShut {NoStop}%
\bibitem [{\citenamefont {Aharonian}\ \emph {et~al.}(2002)\citenamefont
  {Aharonian}, \citenamefont {Belyanin}, \citenamefont {Derishev},
  \citenamefont {Kocharovsky},\ and\ \citenamefont
  {Kocharovsky}}]{Aharonian:2002we}%
  \BibitemOpen
  \bibfield  {author} {\bibinfo {author} {\bibfnamefont {F.~A.}\ \bibnamefont
  {Aharonian}}, \bibinfo {author} {\bibfnamefont {A.~A.}\ \bibnamefont
  {Belyanin}}, \bibinfo {author} {\bibfnamefont {E.~V.}\ \bibnamefont
  {Derishev}}, \bibinfo {author} {\bibfnamefont {V.~V.}\ \bibnamefont
  {Kocharovsky}},\ and\ \bibinfo {author} {\bibfnamefont {V.~V.}\ \bibnamefont
  {Kocharovsky}},\ }\href {https://doi.org/10.1103/PhysRevD.66.023005}
  {\bibfield  {journal} {\bibinfo  {journal} {Phys. Rev. D}\ }\textbf {\bibinfo
  {volume} {66}},\ \bibinfo {pages} {023005} (\bibinfo {year} {2002})},\
  \Eprint {https://arxiv.org/abs/astro-ph/0202229} {arXiv:astro-ph/0202229}
  \BibitemShut {NoStop}%
\bibitem [{\citenamefont {Cao}\ \emph {et~al.}(2021)\citenamefont {Cao} \emph
  {et~al.}}]{LHAASO:2021cbz}%
  \BibitemOpen
  \bibfield  {author} {\bibinfo {author} {\bibfnamefont {Z.}~\bibnamefont
  {Cao}} \emph {et~al.} (\bibinfo {collaboration} {LHAASO*{\textdagger},
  LHAASO}),\ }\href {https://doi.org/10.1126/science.abg5137} {\bibfield
  {journal} {\bibinfo  {journal} {Science}\ }\textbf {\bibinfo {volume}
  {373}},\ \bibinfo {pages} {425} (\bibinfo {year} {2021})},\ \Eprint
  {https://arxiv.org/abs/2111.06545} {arXiv:2111.06545 [astro-ph.HE]}
  \BibitemShut {NoStop}%
\bibitem [{\citenamefont {Nikonorov}\ \emph {et~al.}(2027)\citenamefont
  {Nikonorov}, \citenamefont {Barkov},\ and\ \citenamefont
  {Lyutikov}}]{Nikonorov:2026bvw}%
  \BibitemOpen
  \bibfield  {author} {\bibinfo {author} {\bibfnamefont {I.~N.}\ \bibnamefont
  {Nikonorov}}, \bibinfo {author} {\bibfnamefont {M.~V.}\ \bibnamefont
  {Barkov}},\ and\ \bibinfo {author} {\bibfnamefont {M.}~\bibnamefont
  {Lyutikov}},\ }\href {https://doi.org/10.1016/j.jheap.2026.100710} {\bibfield
   {journal} {\bibinfo  {journal} {JHEAp}\ }\textbf {\bibinfo {volume} {55}},\
  \bibinfo {pages} {100710} (\bibinfo {year} {2027})},\ \Eprint
  {https://arxiv.org/abs/2605.13697} {arXiv:2605.13697 [astro-ph.HE]}
  \BibitemShut {NoStop}%
\bibitem [{\citenamefont {Amenomori}\ \emph {et~al.}(2026)\citenamefont
  {Amenomori} \emph {et~al.}}]{TibetASg:2026bag}%
  \BibitemOpen
  \bibfield  {author} {\bibinfo {author} {\bibfnamefont {M.}~\bibnamefont
  {Amenomori}} \emph {et~al.} (\bibinfo {collaboration} {Tibet
  AS{\ensuremath{\gamma}}*}),\ }\href {https://doi.org/10.1126/sciadv.adv8173}
  {\bibfield  {journal} {\bibinfo  {journal} {Sci. Adv.}\ }\textbf {\bibinfo
  {volume} {12}},\ \bibinfo {pages} {adv8173} (\bibinfo {year}
  {2026})}\BibitemShut {NoStop}%
\bibitem [{\citenamefont {Migliazzo}\ \emph {et~al.}(2002)\citenamefont
  {Migliazzo}, \citenamefont {Gaensler}, \citenamefont {Backer}, \citenamefont
  {Stappers}, \citenamefont {van~der Swaluw},\ and\ \citenamefont
  {Strom}}]{Migliazzo:2002vd}%
  \BibitemOpen
  \bibfield  {author} {\bibinfo {author} {\bibfnamefont {J.~M.}\ \bibnamefont
  {Migliazzo}}, \bibinfo {author} {\bibfnamefont {B.~M.}\ \bibnamefont
  {Gaensler}}, \bibinfo {author} {\bibfnamefont {D.~C.}\ \bibnamefont
  {Backer}}, \bibinfo {author} {\bibfnamefont {B.~W.}\ \bibnamefont
  {Stappers}}, \bibinfo {author} {\bibfnamefont {E.}~\bibnamefont {van~der
  Swaluw}},\ and\ \bibinfo {author} {\bibfnamefont {R.~G.}\ \bibnamefont
  {Strom}},\ }\href {https://doi.org/10.1086/340002} {\bibfield  {journal}
  {\bibinfo  {journal} {Astrophys. J. Lett.}\ }\textbf {\bibinfo {volume}
  {567}},\ \bibinfo {pages} {L141} (\bibinfo {year} {2002})},\ \Eprint
  {https://arxiv.org/abs/astro-ph/0202063} {arXiv:astro-ph/0202063}
  \BibitemShut {NoStop}%
\bibitem [{\citenamefont {Suzuki}\ \emph {et~al.}(2021)\citenamefont {Suzuki},
  \citenamefont {Bamba},\ and\ \citenamefont {Shibata}}]{Suzuki:2021ium}%
  \BibitemOpen
  \bibfield  {author} {\bibinfo {author} {\bibfnamefont {H.}~\bibnamefont
  {Suzuki}}, \bibinfo {author} {\bibfnamefont {A.}~\bibnamefont {Bamba}},\ and\
  \bibinfo {author} {\bibfnamefont {S.}~\bibnamefont {Shibata}},\ }\href
  {https://doi.org/10.3847/1538-4357/abfb02} {\bibfield  {journal} {\bibinfo
  {journal} {Astrophys. J.}\ }\textbf {\bibinfo {volume} {914}},\ \bibinfo
  {pages} {103} (\bibinfo {year} {2021})},\ \Eprint
  {https://arxiv.org/abs/2104.10052} {arXiv:2104.10052 [astro-ph.HE]}
  \BibitemShut {NoStop}%
\bibitem [{\citenamefont {Di~Mauro}\ \emph {et~al.}(2020)\citenamefont
  {Di~Mauro}, \citenamefont {Manconi},\ and\ \citenamefont
  {Donato}}]{DiMauro:2019hwn}%
  \BibitemOpen
  \bibfield  {author} {\bibinfo {author} {\bibfnamefont {M.}~\bibnamefont
  {Di~Mauro}}, \bibinfo {author} {\bibfnamefont {S.}~\bibnamefont {Manconi}},\
  and\ \bibinfo {author} {\bibfnamefont {F.}~\bibnamefont {Donato}},\ }\href
  {https://doi.org/10.1103/PhysRevD.101.103035} {\bibfield  {journal} {\bibinfo
   {journal} {Phys. Rev. D}\ }\textbf {\bibinfo {volume} {101}},\ \bibinfo
  {pages} {103035} (\bibinfo {year} {2020})},\ \Eprint
  {https://arxiv.org/abs/1908.03216} {arXiv:1908.03216 [astro-ph.HE]}
  \BibitemShut {NoStop}%
\bibitem [{\citenamefont {Pan}\ \emph {et~al.}(2024)\citenamefont {Pan},
  \citenamefont {Jiang}, \citenamefont {Yue}, \citenamefont {Lei},
  \citenamefont {Cui},\ and\ \citenamefont {Yuan}}]{Pan:2024adp}%
  \BibitemOpen
  \bibfield  {author} {\bibinfo {author} {\bibfnamefont {X.}~\bibnamefont
  {Pan}}, \bibinfo {author} {\bibfnamefont {W.}~\bibnamefont {Jiang}}, \bibinfo
  {author} {\bibfnamefont {C.}~\bibnamefont {Yue}}, \bibinfo {author}
  {\bibfnamefont {S.-J.}\ \bibnamefont {Lei}}, \bibinfo {author} {\bibfnamefont
  {Y.-X.}\ \bibnamefont {Cui}},\ and\ \bibinfo {author} {\bibfnamefont
  {Q.}~\bibnamefont {Yuan}},\ }\href
  {https://doi.org/10.1007/s41365-024-01499-x} {\bibfield  {journal} {\bibinfo
  {journal} {Nucl. Sci. Tech.}\ }\textbf {\bibinfo {volume} {35}},\ \bibinfo
  {pages} {149} (\bibinfo {year} {2024})},\ \Eprint
  {https://arxiv.org/abs/2407.16973} {arXiv:2407.16973 [astro-ph.IM]}
  \BibitemShut {NoStop}%
\bibitem [{\citenamefont {Qiao}\ \emph {et~al.}(2026)\citenamefont {Qiao},
  \citenamefont {Yuan},\ and\ \citenamefont {Guo}}]{Qiao:2025zyk}%
  \BibitemOpen
  \bibfield  {author} {\bibinfo {author} {\bibfnamefont {B.-Q.}\ \bibnamefont
  {Qiao}}, \bibinfo {author} {\bibfnamefont {Q.}~\bibnamefont {Yuan}},\ and\
  \bibinfo {author} {\bibfnamefont {Y.-Q.}\ \bibnamefont {Guo}},\ }\href
  {https://doi.org/10.1088/1475-7516/2026/07/050} {\bibfield  {journal}
  {\bibinfo  {journal} {JCAP}\ }\textbf {\bibinfo {volume} {07}},\ \bibinfo
  {pages} {050}},\ \Eprint {https://arxiv.org/abs/2506.18118} {arXiv:2506.18118
  [astro-ph.HE]} \BibitemShut {NoStop}%
\bibitem [{\citenamefont {Fang}\ \emph {et~al.}(2020)\citenamefont {Fang},
  \citenamefont {Bi},\ and\ \citenamefont {Yin}}]{Fang:2020cru}%
  \BibitemOpen
  \bibfield  {author} {\bibinfo {author} {\bibfnamefont {K.}~\bibnamefont
  {Fang}}, \bibinfo {author} {\bibfnamefont {X.-J.}\ \bibnamefont {Bi}},\ and\
  \bibinfo {author} {\bibfnamefont {P.-F.}\ \bibnamefont {Yin}},\ }\href
  {https://doi.org/10.3847/1538-4357/abb8d7} {\bibfield  {journal} {\bibinfo
  {journal} {Astrophys. J.}\ }\textbf {\bibinfo {volume} {903}},\ \bibinfo
  {pages} {69} (\bibinfo {year} {2020})},\ \Eprint
  {https://arxiv.org/abs/2003.13635} {arXiv:2003.13635 [astro-ph.HE]}
  \BibitemShut {NoStop}%
\bibitem [{\citenamefont {Blondin}\ \emph {et~al.}(2001)\citenamefont
  {Blondin}, \citenamefont {Chevalier},\ and\ \citenamefont
  {Frierson}}]{Blondin:2001bf}%
  \BibitemOpen
  \bibfield  {author} {\bibinfo {author} {\bibfnamefont {J.~M.}\ \bibnamefont
  {Blondin}}, \bibinfo {author} {\bibfnamefont {R.~A.}\ \bibnamefont
  {Chevalier}},\ and\ \bibinfo {author} {\bibfnamefont {D.~M.}\ \bibnamefont
  {Frierson}},\ }\href {https://doi.org/10.1086/324042} {\bibfield  {journal}
  {\bibinfo  {journal} {Astrophys. J.}\ }\textbf {\bibinfo {volume} {563}},\
  \bibinfo {pages} {806} (\bibinfo {year} {2001})},\ \Eprint
  {https://arxiv.org/abs/astro-ph/0107076} {arXiv:astro-ph/0107076}
  \BibitemShut {NoStop}%
\bibitem [{\citenamefont {{van der Swaluw}}\ \emph {et~al.}(2001)\citenamefont
  {{van der Swaluw}}, \citenamefont {{Achterberg}}, \citenamefont {{Gallant}},\
  and\ \citenamefont {{T{\'o}th}}}]{2001A&A...380..309V}%
  \BibitemOpen
  \bibfield  {author} {\bibinfo {author} {\bibfnamefont {E.}~\bibnamefont {{van
  der Swaluw}}}, \bibinfo {author} {\bibfnamefont {A.}~\bibnamefont
  {{Achterberg}}}, \bibinfo {author} {\bibfnamefont {Y.~A.}\ \bibnamefont
  {{Gallant}}},\ and\ \bibinfo {author} {\bibfnamefont {G.}~\bibnamefont
  {{T{\'o}th}}},\ }\href {https://doi.org/10.1051/0004-6361:20011437}
  {\bibfield  {journal} {\bibinfo  {journal} {Astron. Astrophys.}\ }\textbf
  {\bibinfo {volume} {380}},\ \bibinfo {pages} {309} (\bibinfo {year}
  {2001})}\BibitemShut {NoStop}%
\bibitem [{\citenamefont {Hinton}\ \emph {et~al.}(2011)\citenamefont {Hinton},
  \citenamefont {Funk}, \citenamefont {Parsons},\ and\ \citenamefont
  {Ohm}}]{Hinton:2011ad}%
  \BibitemOpen
  \bibfield  {author} {\bibinfo {author} {\bibfnamefont {J.}~\bibnamefont
  {Hinton}}, \bibinfo {author} {\bibfnamefont {S.}~\bibnamefont {Funk}},
  \bibinfo {author} {\bibfnamefont {R.~D.}\ \bibnamefont {Parsons}},\ and\
  \bibinfo {author} {\bibfnamefont {S.}~\bibnamefont {Ohm}},\ }\href
  {https://doi.org/10.1088/2041-8205/743/1/L7} {\bibfield  {journal} {\bibinfo
  {journal} {Astrophys. J. Lett.}\ }\textbf {\bibinfo {volume} {743}},\
  \bibinfo {pages} {L7} (\bibinfo {year} {2011})},\ \Eprint
  {https://arxiv.org/abs/1111.2036} {arXiv:1111.2036 [astro-ph.HE]}
  \BibitemShut {NoStop}%
\bibitem [{\citenamefont {Knies}\ \emph {et~al.}(2024)\citenamefont {Knies},
  \citenamefont {Sasaki}, \citenamefont {Becker}, \citenamefont {Liu},
  \citenamefont {Ponti},\ and\ \citenamefont {Plucinsky}}]{Knies:2024spk}%
  \BibitemOpen
  \bibfield  {author} {\bibinfo {author} {\bibfnamefont {J.~R.}\ \bibnamefont
  {Knies}}, \bibinfo {author} {\bibfnamefont {M.}~\bibnamefont {Sasaki}},
  \bibinfo {author} {\bibfnamefont {W.}~\bibnamefont {Becker}}, \bibinfo
  {author} {\bibfnamefont {T.}~\bibnamefont {Liu}}, \bibinfo {author}
  {\bibfnamefont {G.}~\bibnamefont {Ponti}},\ and\ \bibinfo {author}
  {\bibfnamefont {P.~P.}\ \bibnamefont {Plucinsky}},\ }\href
  {https://doi.org/10.1051/0004-6361/202348834} {\bibfield  {journal} {\bibinfo
   {journal} {Astron. Astrophys.}\ }\textbf {\bibinfo {volume} {688}},\
  \bibinfo {pages} {A90} (\bibinfo {year} {2024})},\ \Eprint
  {https://arxiv.org/abs/2401.17289} {arXiv:2401.17289 [astro-ph.HE]}
  \BibitemShut {NoStop}%
\bibitem [{\citenamefont {{Lyne}}\ \emph {et~al.}(1996)\citenamefont {{Lyne}},
  \citenamefont {{Pritchard}}, \citenamefont {{Graham-Smith}},\ and\
  \citenamefont {{Camilo}}}]{1996Natur.381..497L}%
  \BibitemOpen
  \bibfield  {author} {\bibinfo {author} {\bibfnamefont {A.~G.}\ \bibnamefont
  {{Lyne}}}, \bibinfo {author} {\bibfnamefont {R.~S.}\ \bibnamefont
  {{Pritchard}}}, \bibinfo {author} {\bibfnamefont {F.}~\bibnamefont
  {{Graham-Smith}}},\ and\ \bibinfo {author} {\bibfnamefont {F.}~\bibnamefont
  {{Camilo}}},\ }\href {https://doi.org/10.1038/381497a0} {\bibfield  {journal}
  {\bibinfo  {journal} {\nat}\ }\textbf {\bibinfo {volume} {381}},\ \bibinfo
  {pages} {497} (\bibinfo {year} {1996})}\BibitemShut {NoStop}%
\bibitem [{\citenamefont {Espinoza}\ \emph {et~al.}(2017)\citenamefont
  {Espinoza}, \citenamefont {Lyne},\ and\ \citenamefont
  {Stappers}}]{Espinoza:2016ars}%
  \BibitemOpen
  \bibfield  {author} {\bibinfo {author} {\bibfnamefont {C.~M.}\ \bibnamefont
  {Espinoza}}, \bibinfo {author} {\bibfnamefont {A.~G.}\ \bibnamefont {Lyne}},\
  and\ \bibinfo {author} {\bibfnamefont {B.~W.}\ \bibnamefont {Stappers}},\
  }\href {https://doi.org/10.1093/mnras/stw3081} {\bibfield  {journal}
  {\bibinfo  {journal} {Mon. Not. Roy. Astron. Soc.}\ }\textbf {\bibinfo
  {volume} {466}},\ \bibinfo {pages} {147} (\bibinfo {year} {2017})},\ \Eprint
  {https://arxiv.org/abs/1611.08314} {arXiv:1611.08314 [astro-ph.HE]}
  \BibitemShut {NoStop}%
\bibitem [{\citenamefont {{Alford}}\ \emph {et~al.}(2025)\citenamefont
  {{Alford}}, \citenamefont {{Zhang}},\ and\ \citenamefont
  {{Gelfand}}}]{2025ApJ...987...63A}%
  \BibitemOpen
  \bibfield  {author} {\bibinfo {author} {\bibfnamefont {J.~A.~J.}\
  \bibnamefont {{Alford}}}, \bibinfo {author} {\bibfnamefont {G.-B.}\
  \bibnamefont {{Zhang}}},\ and\ \bibinfo {author} {\bibfnamefont {J.~D.}\
  \bibnamefont {{Gelfand}}},\ }\href {https://doi.org/10.3847/1538-4357/add92e}
  {\bibfield  {journal} {\bibinfo  {journal} {\apj}\ }\textbf {\bibinfo
  {volume} {987}},\ \bibinfo {eid} {63} (\bibinfo {year} {2025})},\ \Eprint
  {https://arxiv.org/abs/2506.22409} {arXiv:2506.22409 [astro-ph.HE]}
  \BibitemShut {NoStop}%
\bibitem [{\citenamefont {Evoli}\ \emph {et~al.}(2018)\citenamefont {Evoli},
  \citenamefont {Linden},\ and\ \citenamefont {Morlino}}]{Evoli:2018aza}%
  \BibitemOpen
  \bibfield  {author} {\bibinfo {author} {\bibfnamefont {C.}~\bibnamefont
  {Evoli}}, \bibinfo {author} {\bibfnamefont {T.}~\bibnamefont {Linden}},\ and\
  \bibinfo {author} {\bibfnamefont {G.}~\bibnamefont {Morlino}},\ }\href
  {https://doi.org/10.1103/PhysRevD.98.063017} {\bibfield  {journal} {\bibinfo
  {journal} {Phys. Rev. D}\ }\textbf {\bibinfo {volume} {98}},\ \bibinfo
  {pages} {063017} (\bibinfo {year} {2018})},\ \Eprint
  {https://arxiv.org/abs/1807.09263} {arXiv:1807.09263 [astro-ph.HE]}
  \BibitemShut {NoStop}%
\bibitem [{\citenamefont {Fang}\ \emph {et~al.}(2019)\citenamefont {Fang},
  \citenamefont {Bi},\ and\ \citenamefont {Yin}}]{Kun:2019sks}%
  \BibitemOpen
  \bibfield  {author} {\bibinfo {author} {\bibfnamefont {K.}~\bibnamefont
  {Fang}}, \bibinfo {author} {\bibfnamefont {X.-J.}\ \bibnamefont {Bi}},\ and\
  \bibinfo {author} {\bibfnamefont {P.-F.}\ \bibnamefont {Yin}},\ }\href
  {https://doi.org/10.1093/mnras/stz1974} {\bibfield  {journal} {\bibinfo
  {journal} {Mon. Not. Roy. Astron. Soc.}\ }\textbf {\bibinfo {volume} {488}},\
  \bibinfo {pages} {4074} (\bibinfo {year} {2019})},\ \Eprint
  {https://arxiv.org/abs/1903.06421} {arXiv:1903.06421 [astro-ph.HE]}
  \BibitemShut {NoStop}%
\end{thebibliography}%

\end{document}